\documentclass{ws-rv961x669}
\usepackage{ws-rv-van}     % numbered citation/references (default)
\usepackage{ws-rv-thm}     % comment this line when `amsthm / theorem / ntheorem` package is used
\usepackage{aas_macro}
\makeindex
\begin{document}
%\title{Neutrinos from Gamma-ray Bursts}
%\chapter[Neutrinos from GRBs]{Neutrinos from Gamma-ray Bursts}\footnote{To be published in Neutrino Physics and Astrophysics, edited by F. W. Stecker, in the Encyclopedia of Cosmology II, edited by G. G. Fazio, World Scientific Publishing Company, Singapore, 2022}
\begin{center}
{\bf Neutrinos from Gamma-ray Bursts\footnote{To be published in Neutrino Physics and Astrophysics, edited by F. W. Stecker, in the Encyclopedia of Cosmology II, edited by G. G. Fazio, World Scientific Publishing Company, Singapore, 2022}}
 
\end{center}

\author[S.S. Kimura]{Shigeo S. Kimura}%\footnote{Author footnote.}}
%\index[aindx]{Author, F.} % or \aindx{Author, F.}
%\index[aindx]{Author, S.} % or \aindx{Author, S.}

\address{Frontier Research Institute for Interdisciplinary Sciences,  \\
Tohoku University, Sendai 980-8578, Japan,\\
shigeo@astr.tohoku.ac.jp}

\begin{abstract}
Gamma-ray bursts (GRBs) are the most luminous electromagnetic burst in the Universe. They occur when a rapidly rotating massive star collapses or a binary neutron star merges. These events leave a newborn central compact object, either a black hole or neutron star, which launches relativistic jets that emit the luminous gamma-ray signals. These jets can accelerate non-thermal protons, which are expected to produce high-energy neutrinos via photohadronic interactions. 
%At the central part of the explosion, there is an accretion disk with an enormous accretion rate, which emit MeV neutrinos through thermal processes. 
This Chapter briefly summarizes the current physical picture of GRBs, and discusses neutrino emissions from GRBs, including both prompt and afterglow phases. Neutrinos from sub-classes of GRBs, including low-luminosity GRBs and short GRBs, are also discussed. 
\end{abstract}

%\markboth{Even Page Header}{Odd Page Header} % Customized running heads

\body

\tableofcontents

\section{Introduction}

A gamma-ray burst (GRB) is a bright transient event in the soft gamma-ray band, which is followed by an afterglow, a broadband (radio/optical/X-ray/GeV-TeV gamma-ray) transient with a longer timescale of hours to days. The typical apparent luminosity and duration for GRBs are $10^{52}\rm~erg~s^{-1}$ and 30 seconds, respectively. This is the most luminous explosion in electromagnetic waves, and the typical apparent total radiation energy of gamma rays reaches $3\times10^{53}$ erg. Our understandings of GRBs have been drastically improved since the discovery in 1967. Here, we briefly review the history of GRB observations (see e.g., Ref. \cite{2018pgrb.book.....Z} for the detailed review of GRB history).

The first GRB was discovered by a military satellite system {\it Vela}~\cite{1973ApJ...182L..85K}, which aimed to monitor nuclear bomb experiments from space. This system found bright gamma-ray signals from neither the Earth nor the Sun, posing a big mystery in astrophysics. In 1990s, Burst And Transient Source Experiment (BATSE) found several important features of GRBs, including the isotropic distribution in the sky that ruled out their galactic origin\cite{Meegan:1992xg}, broken power-law spectra that suggest non-thermal particle production\cite{BMF93a}, the bi-modal distribution of the burst duration which hints two sub-classes in GRBs (long and short GRBs; LGRBs and SGRBs, respectively)\cite{1993ApJ...413L.101K}. 
%During this era, there was an intense debate whether GRBs are events occurring near the solar system or at a cosmological distance.

In 1997, Beppo-SAX satellite observed the first X-ray counterpart from GRB 970228\cite{1997Natur.387..783C}, from which optical counterpart was also detected\cite{Sahu:1997vp,1997Natur.386..686V}. Beppo-SAX also detected GRB 970508, from which radio, optical, and X-ray counterparts were detected\cite{1997Natur.389..261F}. This is the beginning of the multi-wavelength afterglow era. The multi-wavelength information enables us to identify the host galaxy of GRB 970508, which turned out that the GRB occured in a cosmological distance with its redshift $z=0.835$. With this redshift, the total energy of the GRB exceeds $10^{50}$ erg, revealing that GRBs are the most energetic explosion in electromagnetic waves. Next year, Beppo-SAX detected GRB 980425 that is accompanied with a luminous broad-lined type-Ic supernova (SN), SN1998bw\cite{1998Natur.395..663K}. This confirms that at least some fraction of GRBs occur when massive stars collapse. This scenario is known as the collapsar model\cite{1993ApJ...405..273W,1998ApJ...494L..45P}. The luminosity of GRB 980425 is much lower than those of typical GRBs, and thus,this is the discovery of the first low-luminosity GRB (LLGRB). In 2003, HETE-2 satellite detected a luminous GRB accompanied with a luminous type-Ic SN, SN2003dh \cite{2003ApJ...591L..17S}. This GRB-SN connection strengthened the collapsar model, and roughly 50 associations are reported by 2017\cite{2017AdAst2017E...5C}.

In 2004, Niel-Gherels Swift Observatory was launched, which aimed to quickly follow-up GRBs. X-ray Telescope (XRT) onboard Swift detected the first X-ray afterglow from an SGRB, GRB 050509B \cite{2005Natur.437..851G}, as well as GRB 050709 and GRB 050724 shortly after it \cite{2005Natur.437..859H,2005Natur.438..994B}. These bursts are localized to elliptical galaxies at redshift $z\sim0.2$, confirming that the SGRBs are extragalactic events. Also, Swift observed the early phase of the afterglows within an hour after the prompt bursts for both LGRBs and SGRBs, which revealed that the central engine is active for at least minutes to hours after the prompt burst for both LGRBs and SGRBs\cite{2005Sci...309.1833B,NB06a,Swift11a}. 

In 2008, Gamma-ray Burst Monitor (GBM) and Large Area Telescope (LAT) onboard Fermi satellite were started in operation. The field of view for GBM covers almost all the sky and a wide range of gamma-ray energy, from $\sim10$ keV to $\sim25$ MeV. These advantages increase the statistics of the GRBs\cite{2016ApJS..223...28N}. LAT detected mysterious high-energy gamma-rays ($>100$ MeV) from some fraction of bright LGRBs and SGRBs\cite{Ajello:2019zki}. Regarding the high-energy gamma-ray, MAGIC and HESS reported detection of very-high-energy gamma-rays ($>100$ GeV) from GRB 190114C and GRB 180720B in 2019\cite{Acciari:2019dbx,Arakawa:2019cfc}. The high-energy gamma-ray observations will enable us to narrow down the nature of the afterglow microphysics, such as the magnetic field amplifications and non-thermal electron production efficiency. 

In 2010s, GRB physics entered the multi-messenger era. The IceCube Collaboration reported non-detection of high-energy neutrinos from gamma-ray detected GRBs, which put a meaningful constraint on the physical quantities of GRB jets\cite{IceCube17b}.
In 2017, the LIGO/VIRGO collaboration reported detection of gravitational-wave signal from a binary neutron-star (BNS) merger, and 2 second later, Fermi-GBM observed a short gamma-ray burst\cite{LIGO17d}. This is the first firmly identified electromagnetic counterpart to the gravitational wave event. Also, this is the first detection of low-luminosity SGRB. Based on the observations and modelings of afterglow emission\citep{MDG18a,2018MNRAS.478L..18T}, this event is likely a canonical SGRB seen from an off-axis observer, strongly supporting the BNS merger model as the origin of the SGRBs\cite{1986ApJ...308L..43P,ELP89a}.

High-energy neutrinos can be produced in both of the prompt and afterglow phases. Also, all the sub-classes of GRBs, i.e., LGRBs, SGRBs, LLGRBs, and failed GRBs (or choked GRBs) can be neutrino sources. In this chapter, we first briefly review the basic concepts of GRBs in Section \ref{sec:basic}. Then, we will discuss the neutrino emissions from the prompt bursts in Section \ref{sec:prompt}, the afterglow phase in Section \ref{sec:afterglow}, the LLGRBs and choked GRBs in Section \ref{sec:llgrb}, and SGRBs in Section \ref{sec:bns}.
%and the central engine (Section \ref{sec:engine}). 
%In each Section, observations and basic physical pictures are summarized.

\section{Basic Concepts of GRBs} \label{sec:basic}

\begin{figure}[tb]
 \begin{center}
  \includegraphics[width=\linewidth]{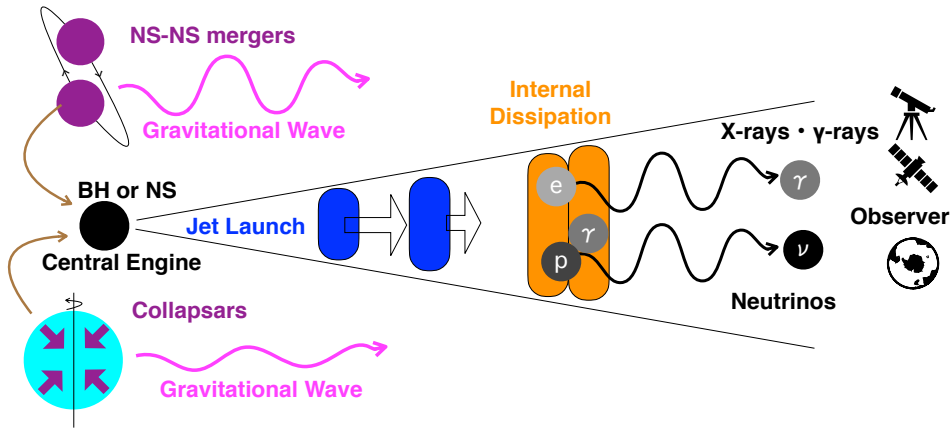}
  \caption{Schematic picture of the prompt bursts of GRBs. A collapsar, a collapsing rapidly rotating Wolf-Rayet star, or a merging BNS creates a remnant compact object, either a rapidly rotating black hole or a highly magnetized neutron star (NS). The remnant object launches relativistic jets, which dissipate their kinetic energies through shocks formed by collisions of the jets. These shocks accelerate non-thermal electrons and protons. The electrons emit observed gamma-rays through the synchrotron process. The protons interact with the gamma-rays, which lead to high-energy neutrino production via the photo-meson production process. }\label{fig:prompt}
 \end{center}
\end{figure}

Owing to the improvements of the observational techniques and theoretical modelings, a standard physical picture of GRBs is constructed.  Figure \ref{fig:prompt} shows the schematic picture of the GRB prompt emission. A collapsar or a BNS merger creates a central compact object, which launches relativistic jets. Non-thermal electrons are accelerated through dissipation of the kinetic energy of the jets, leading to gamma-ray emissions observed as GRBs. Although this picture still has a few shortcomings, it can explain several aspects of GRBs.  Before describing the neutrino emissions from GRBs, we overview the basic pictures of GRBs in this section.

\subsection{Observational features}

\begin{figure}[tb]
 \begin{center}
  \includegraphics[width=\linewidth]{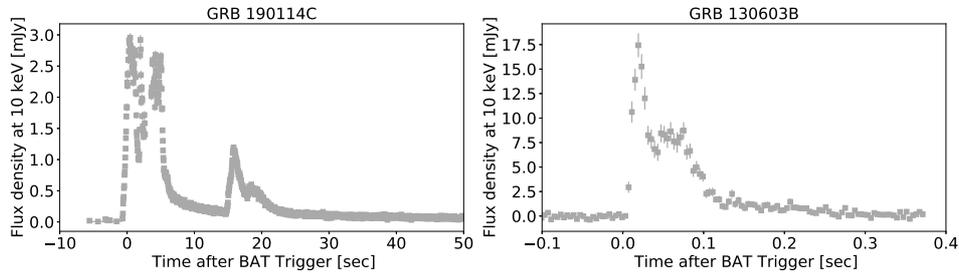}
  \caption{Examples of lightcurves of a long GRB (GRB 190114C; left) and a short GRB (GRB 130603B; right). The data are taken from the Burst Analyzer \cite{2010A&A...519A.102E}. }\label{fig:bat_lightcurve}
 \end{center}
\end{figure}

Prompt phases of GRBs indicate a variety of lightcurves (see Figure \ref{fig:bat_lightcurve}). Some show bumpy lightcurves with a strong variability in a short timescale, while others have smooth lightcurves without any short-timescale structure. The shortest variability timescale is around $10^{-3}$ sec, which is about a dynamical timescale of the innermost stable circular orbit for a 3-$M_\odot$ black hole (BH). The distribution of the duration of the prompt phase, $T_{90}$, indicates a bi-modality, which suggests two sub-classes of GRBs: SGRBs with $T_{90}<2$ sec and LGRBs for $T_{90}>2$ sec. SGRBs typically have $T_{90}\sim0.3$ sec and LGRBs are $T_{90}\sim10$ sec \cite{1993ApJ...413L.101K,2016ApJS..223...28N}. They are highly likely to have different progenitors, which is supported by the observations of SN-GRB\cite{2017AdAst2017E...5C} and GW-GRB \cite{LIGO17e} associations. 

The observed gamma-ray spectra of GRBs can be fitted by a broken power-law function:\cite{BMF93a,2014ApJS..211...12G}
\begin{eqnarray}
 \frac{dN}{dE_\gamma}\propto\left\{
  \begin{array}{ll}
   (E_\gamma/E_{\gamma,\rm br})^{-\alpha_1} ~~&~~ (E_\gamma<E_{\gamma,\rm br}), \\
   (E_\gamma/E_{\gamma,\rm br})^{-\alpha_2} ~~&~~ (E_\gamma>E_{\gamma,\rm br}), 
  \end{array}\right. \label{eq:Band}
\end{eqnarray}
where $E_{\gamma,\rm br}\sim0.1-1$ MeV is the spectral peak energy of prompt gamma-rays and $\alpha_1\sim0-1$ and $\alpha_2\sim2-3$ are the spectral indices for a lower and higher energies, respectively.
Some GRBs show a very hard spectral index, i.e. a low value of $\alpha_1$. A fraction of GRBs have $\alpha<2/3$. This feature cannot be explained by the synchrotron emission by non-thermal electrons\cite{1998ApJ...506L..23P}. Such a hard spectral feature can be interpreted as a photospheric emission that can produce $\alpha_1\sim-1$\cite{1986ApJ...308L..47G}.

For LGRBs, the break energy in the gamma-ray spectrum in the engine frame, $\hat{E}_{\gamma,\rm br}$, correlates with the isotropic equivalent gamma-ray total radiation energy, $\mathcal{E}_{\gamma,\rm iso}$ (Amati relation\cite{2002A&A...390...81A}; see Section \ref{sec:relativistic} for definition of these quantities). A more energetic GRB has a higher break energy. This relation is satisfied for LLGRBs. SGRBs show a similar trend, but have a higher $\hat{E}_{\gamma,\rm br}$ than LGRBs for a fixed $\mathcal{E}_{\gamma,\rm iso}$. $\hat{E}_{\gamma,\rm br}$ also shows a correlation with the gamma-ray peak luminosity, $L_{\gamma,\rm iso}$. A more luminous burst has a higher $\hat{E}_{\gamma,\rm br}$ (Yonetoku relation\cite{2004ApJ...609..935Y}). Both SGRBs and LGRBs lie in the same relation in the $L_{\gamma,\rm iso}-\hat{E}_{\gamma,\rm br}$ plane, while LLGRBs do not follow the same relation. GRB prompt emission models should satisfy these relations, although the origin of the relations are currently unclear.
% We may be able to estimate the distance to the GRB events using these relations\cite{2004MNRAS.348..999I}. However, the dispersion of these relations are too large to accurately estimate the cosmological parameters using GRBs.

\subsection{Compactness problem}

Lightcurves of GRBs exhibit very short variability timescale, $t_{\rm var}\lesssim10$ msec. If the emission region is at rest, such a short variability timescale requires the size of the emission region to be $R\approx t_{\rm var}c\sim3\times10^8t_{\rm var,-2}$ cm\footnote{Hereafter, we use a notation of $Q_X=Q/10^X$ in cgs unit unless otherwise noted.}. The typical luminosity, break gamma-ray energy, and gamma-ray spectral index of a GRB is $\mathcal{E}_{\gamma,\rm iso}\sim10^{52}\rm~erg~s^{-1}$, $E_{\gamma,\rm br}\sim300$ keV, and $\alpha_2\sim2.2$, respectively. Using these values, the differential photon density at the energy $E_\gamma$, $n_\gamma(E_\gamma)=E_\gamma dN/(dE_\gamma dV)$, is estimated to be 
\begin{equation}
n_\gamma(E_\gamma)\sim\frac{3\mathcal{E}_{\gamma,\rm iso}}{4\pi E_{\gamma,\rm br}t_{\rm var}^3c^3}\left(\frac{E_\gamma}{E_{\gamma,\rm br}}\right)^{1-\alpha_2}.  
\end{equation}
These photons interact with photons of $\tilde{E}_\gamma\sim m_e^2c^4/E_\gamma$ through two-photon pair production. The optical depth for $\gamma\gamma$ pair production at the gamma-ray energy $E_\gamma$ is estimated to be 
\begin{eqnarray}
\tau_{\gamma\gamma}^{\rm rest}(E_\gamma)\approx \sigma_{\gamma\gamma}Rn_\gamma(\tilde{E}_\gamma)\approx \frac{3\sigma_{\gamma\gamma}\mathcal{E}_{\gamma,\rm iso}}{4\pi E_{\gamma,\rm br}t_{\rm var}^2c^2}\left(\frac{m_e^2c^4}{E_\gamma E_{\gamma,\rm br}}\right)^{1-\alpha_2}\\
\sim4\times10^{15}t_{\rm var,-2}^{-2}\mathcal{E}_{\gamma,\rm iso,52}\left(\frac{E_{\gamma,\rm br}}{300\rm~keV}\right)^{\alpha_2-2}\left(\frac{E_\gamma}{1\rm~MeV}\right)^{\alpha_2-1},
\end{eqnarray}
where we use $\alpha_2=2.2$ and $\sigma_{\gamma\gamma}\sim0.1\sigma_T$ for the last equation ($\sigma_T$ is the Thomson cross section).
Therefore, the GRBs are too luminous for gamma-rays to escape from the emission region if the emission region is at rest. The gamma-rays are completely absorbed through the pair production processes, and the emergent photon spectrum should be the Planck spectrum with such a high optical depth, which is inconsistent with the observed power-law spectra. This is known as the compactness problem.

\subsection{Relativistic effects} \label{sec:relativistic}

Relativistic motion toward the observer can solve the compactness problem. 
In order to describe the phenomena happening in the relativistically moving fluid at a cosmological distance, we use three inertial frames: the rest frame of the central engine (the engine frame), the rest frame of the fluid (the comoving frame), and the rest frame of the observer (the observer frame). The engine frame and the observer frame are the same inertial frame if we ignore the cosmological expansion. The comoving frame is moving with respect to the two other frames with a Lorentz factor of $\Gamma_j=1/\sqrt{1-\beta_j^2}$, where $\beta_j$ is the fluid velocity in unit of the speed of light. The particle energies of the three frames are related by $E_i=\hat{E}_i/(1+z)$ and $\hat{E}_i=\Gamma_j \varepsilon_i$, where $E_i$, $\hat{E}_i$, and $\varepsilon_i$ are the energy of particle species $i$ in the observer, engine, and comoving frames, respectively, and  $z$ is the redshift.

Let us write the timescales of the observer frame, the engine frame, and the comoving frame as  $T$, $\hat{t}$, and $t$, respectively. We can write the relation among the three timescales as $t = \hat{t}/\Gamma_j$ and $T \approx (1+z) \hat{t}/\Gamma_j^2$. The former relation is the  time dilation by the Lorentz transformation. The latter relation is obtained by effects of the photon propagation and the cosmological redshift. In order to see this, let us consider an expanding shell with velocity $\beta_j c$ that starts emitting photons at $\hat{t}=\hat{t}_1$. We ignore the cosmological redshift for simplicity. The observer at distance $d$ begins to receive the photons at $T=T_1=\hat{t}_1 + d/c$. If the shell stops emitting photons at $\hat{t}=\hat{t}_2=\hat{t}_1+\Delta \hat{t}$, the observer stops receiving photons at  $T=T_2=\hat{t}_1 + \Delta \hat{t} + (d - c\beta_j\Delta \hat{t})/c$. Then, the observed duration of the event is $\Delta T = T_2 - T_1 = \Delta \hat{t} (1 - \beta_j )\approx \Delta \hat{t} /(2 \Gamma_j^2)$. Therefore, the observer timescale is different from the engine timescale even if we ignore the cosmological expansion.

The modifications of the photon energy and variability timescale solve the compactness problem. In order to understand these effects, let us consider a spherically expanding shell again.  The emission radius is modified to be $R \approx c \hat{t}_{\rm var} \approx2 \Gamma_j^2T_{\rm var} c \simeq 6\times10^{13}T_{\rm var,-2}\Gamma_{j,2.5}^2$ cm in the engine frame, where  we ignore the cosmological effect for simplicity (usually, this effect is very minor, compared to the relativistic motion).  The photon energy and total radiation energy in the comoving frame are given by $\varepsilon_\gamma = E_\gamma/\Gamma_j$ and $\mathcal{E}'_{\gamma,\rm iso} = \mathcal{E}_{\gamma,\rm iso}/\Gamma_j$, where $Q'$ denotes the quantities in the comoving frame. Taking these two effects into account, the number density of the photons in the comoving frame is given by 
\begin{equation}
n'_\gamma(\varepsilon_\gamma)\approx\frac{\mathcal{E}'_{\gamma,\rm iso}}{4\pi R^2\Delta R'\varepsilon_{\gamma,\rm br}}\left(\frac{\varepsilon_\gamma}{\varepsilon_{\gamma,\rm br}}\right)^{1-\alpha_2},%\approx\frac{\mathcal{E}_{\gamma,\rm iso}}{16\pi \Gamma_j^4t_{\rm var}^2c^2\Delta R'E_{\gamma,\rm br}}\left(\frac{E_\gamma}{E_{\gamma,\rm br}}\right)^{1-\alpha_2},
\end{equation}
$\Delta R'$ is the width of the shell. Now, the photon energy is Doppler de-boosted, and then, the two-photon pair production occurs with $\tilde{\varepsilon}_\gamma\sim m_e^2c^4/\varepsilon_\gamma\approx \Gamma_jm_e^2c^4/E_\gamma$. The optical depth for a photon of $E_\gamma$ in the observer frame is then estimated to be 
\begin{eqnarray}
\tau_{\gamma\gamma}(E_\gamma)\approx\tau'_{\gamma\gamma}(\varepsilon_\gamma)\approx n'_\gamma(\tilde{\varepsilon}_\gamma)\Delta R'\sigma_{\gamma\gamma}\approx \frac{\sigma_{\gamma\gamma}\mathcal{E}_{\gamma,\rm iso}}{16\pi \Gamma_j^4c^2t_{\rm var}^2E_{\gamma,\rm br}}\left(\frac{\Gamma_j^2m_e^2c^4}{E_\gamma E_{\gamma,\rm br}}\right)^{1-\alpha_2}\\
\sim 0.04\mathcal{E}_{\gamma,\rm iso,52}\Gamma_{j,2.5}^{-2\alpha_2-2}t_{\rm var,-2}^{-2}\left(\frac{E_\gamma}{1\rm~MeV}\right)^{\alpha_2-1}\left(\frac{E_{\gamma,\rm br}}{300\rm~keV}\right)^{\alpha_2-2},\nonumber
\end{eqnarray}
where we use $\alpha_2=2.2$ for the last equation. We can write $\tau_{\gamma\gamma}(E_\gamma)\sim\Gamma_j^{-\alpha_2-2}\tau_{\gamma\gamma}^{\rm rest}(E_\gamma)/12$, from which we see that the relativistic motion reduces the optical depth by $\Gamma_j^{-6}-\Gamma_j^{-7}$ for the range of observed spectral index, $2<\alpha_2<2.5$. Therefore, the compactness problem is solved for the expanding shell of $\Gamma_j\gtrsim200-300$ for typical GRBs.

In reality, the expanding shell is a narrow jet. Since the photons from the jet is beamed toward the angle of $1/\Gamma_j$, we cannot distinguish the spherically expanding shell from the jet as long as the jet opening angle, $\theta_j$, is larger than $1/\Gamma_j$. Thus, we use ``isotropic equivalent'' luminosity and total radiation energy, $L_{\gamma,\rm iso}=4\pi d_L^2 F_\gamma$ and $\mathcal{E}_{\gamma,\rm iso}=4\pi d_L^2 S_\gamma/(1+z)$, where $d_L$ is the luminosity distance, $F_\gamma$ is the observed gamma-ray energy flux, and $S_\gamma$ is the observed gamma-ray energy fluence\footnote{Here, we ignore the observed band width for simplicity. The correction from the observed band to the bolometric one can have a significant impact on the total energy of GRBs.}. Their relation to the intrinsic jet luminosity and total radiation energy are written as $L_{\gamma,j}\approx \theta_j^2 L_{\gamma,\rm iso}/2$ and $\mathcal{E}_{\gamma,j}\approx \theta_j^2 \mathcal{E}_{\gamma,\rm iso}/2$, respectively.

\subsection{Central engine of the GRB jets}

The launching mechanism of the relativistic jets is a long-standing problem in astrophysics.  Both of the progenitor systems of LGRBs and SGRBs, namely collapsars and BNS mergers, lead to a central compact object surrounded by an accretion flow because of the angular momentum distribution of the system. 

The accretion rate onto the compact object is so high that it can power the energetic explosion.
Regarding the collapsar scenario, the progenitor star is rapidly rotating, and the outer region of the progenitor star cannot directly collapse into a BH because of its high angular momentum. Based on the stellar angular momentum distribution right before the collapse \cite{WH06a}, the stellar material at $R_*\gtrsim(0.5-3)\times10^9$ cm has an angular momentum enough to form an accretion disk around the central BH, and enclosed mass within it is about $M_*\sim2-5~M_\odot$ \cite{ZMK18a}. Then, the mass accretion rate onto the central object at the time of the disk formation is estimated to be $\dot{M}\sim M_*/t_{\rm fall}\sim 2 M_{*,3}^{3/2}R_{*,9}^{-3/2}~M_\odot\rm~s^{-1}$, where $t_{\rm fall}=\sqrt{R_*^3/(GM_*)}$ is the free fall time and $M_{*,X}=(M_*/X~M_\odot)$. 
In terms of the BNS merger scenario, a lot of numerical relativity simulations on BNS mergers are performed (see Refs. \cite{2017RPPh...80i6901B,2019ARNPS..69...41S} for recent reviews). According to the simulation results, the remnant central object with $M=M_*$ is surrounded by a massive torus whose mass and radius is $M_{\rm trs}\sim0.1~M_\odot$ and $R_{\rm trs}\sim10^7$ cm, respectively (see e.g., initial conditions of Refs. \cite{SM17a,2018ApJ...860...64F}). The accretion timescale of the torus is estimated to be~\cite{pri81} $t_{\rm vis}\sim \alpha^{-1} h^{-2} \sqrt{R_{\rm trs}^3/(GM_*)}\sim0.2{\rm~s~}\alpha_{-1}h_{-0.5}^{-2} M_{*,2.5}^{-1/2}R_{\rm trs,7}^{3/2}$, where $\alpha\sim0.1$ is the viscous parameter\cite{ss73} and $h$ is the aspect ratio of the accretion torus. Then, the mass accretion rate is estimated to be $\dot{M}\sim M_{\rm trs}/t_{\rm vis}\sim0.6~M_\odot~{\rm s^{-1}}\alpha_{-1}h_{0.5}^2M_{*,2.5}^{1/2}R_{\rm trs,7}^{-3/2}M_{\rm trs,-1}$, where $M_{\rm trs,-1}=M_{\rm trs}/(0.1~M_\odot)$.
The accretion luminosity is estimated to be $\dot{M}c^2\sim2\times10^{54}{\rm~erg~s^{-1}}(\dot{M}/1 M_\odot\rm~s^{-1})$, which is sufficient to power the observed GRBs for both situations.

With such a high accretion rate, photons are completely trapped in the accretion flow. The flow is so hot that MeV neutrinos can be produced by thermal processes, forming a neutrino-cooling dominated accretion flows (NDAFs\cite{PWF99a,2001ApJ...557..949N,KM02a}). Both neutrinos and anti-neutrinos are produced from the accretion flows, and they interact at the polar region of the BH. This mechanism efficiently produces electron-positron pairs and injects a large amount of energy into the funnel. 
With a higher spin BH, the innermost stable circular orbit (ISCO) is smaller, leading to a higher neutrino luminosity. A higher mass accretion rate also makes the neutrino luminosity higher. Thus, the energy injection rate provided by the neutrino pair annihilation mechanism is higher for a high spin BH (described by the dimensionless spin parameter $a$) and a high mass accretion rate, $\dot{M}$. For a highly spining BH of $a=0.95$, the energy injection rate can be as high as $L_{\nu\overline{\nu}}\sim10^{52}\rm~erg~s^{-1}$ with $\dot{M}\sim1~M_\odot~\rm s^{-1}$ \cite{2013ApJ...765..125L,2017ApJ...849...47L}. This is sufficient to power most of the observed GRBs.

The thermal neutrino luminosity from an NDAF is very high, $L_\nu\sim0.1\dot{M}c^2 \sim10^{53}\rm~erg~s^{-1}$, which is comparable to that from a core-collapse SN. However, we cannot expect neutrino detection from an NDAF because of the event rate of GRBs. The detection horizon of the neutrinos from a typical SN is at most several Mpc even with future experiments, such as Hyper-Kamiokande \cite{Abe:2018uyc}.  The intrinsic event rate of GRBs are estimated to be a thousand times lower than that of typical core-collapse SNe, and thus, it is very unlikely for a GRB to occur within a detection horizon. Several experiments searched for MeV neutrinos from GRBs, but no neutrinos were found so far\cite{Asakura:2015fua,Abe:2020zpn,Abe:2018uyc}. This is consistent with the theoretical expectations\cite{2018PhRvD..97j3001K}.

Blandford-Znajek mechanism is also actively discussed as an alternative for the energy injection \cite{BZ77a,Kom04a,TT16a}. The rotation energy of the central spinning BH can be extracted as a Poyntyng flux if a strong and ordered magnetic field threads the spinning BH. This mechanism also supplies sufficient power to the jets.

\subsection{Fireball and baryonic jets} \label{sec:fireball}

Although the production mechanism of the relativistic jets is still under debate, we here discuss the hot fireball model. This is the mainstream model of the GRB phenomenology. In this model, we assume that a large amount of thermal energy is concentrated on a small region ($L_{\rm iso}\sim10^{51}-10^{53}\rm~erg~s^{-1}$ and $R_{\rm ini}\sim10^7$ cm) via the mechanisms mentioned in the previous subsection. This ``fireball'' expands using its thermal pressure, efficiently converting its thermal energy to its kinetic energy. Here, we discuss the evolution of the fireball assuming the spherical symmetry. We can regard a part of the spherical shell as the jets. 

Initially, the fireball is not moving with a highly relativistic speed. The initial temperature of the fireball is roughly estimated to be $T_{\rm ini}\sim L_{\rm fb}^{1/4}/(4\pi R_{\rm ini}^2ca_r)^{1/4}\sim1{\rm~MeV}L_{\rm fb,52}^{1/4}R_{\rm ini,7}^{-1/2}$\footnote{In this subsection, we use $T$ as the temperature, instead of the observer time. }, where $a_r$ is the radiation constant and we assume the thermal energy, $U_{\rm ini}$, is dominated by radiation and relativistic particles, $U_{\rm ini}\approx a_rT_{\rm ini}^4$. During the fireball expansion, the width of the expanding shell remains almost constant in the engine frame, $\Delta_{\rm sh}\approx R_{\rm ini}$ because both of the front and back ends of the shell move with the speed of light. In contrast, the shell width in the comoving frame increases with its Lorentz factor as $\Delta'_{\rm sh}\approx \Gamma R_{\rm ini}$. The volume of the fireball shell is then given by $V'\approx 4\pi R^2\Gamma R_{\rm ini}$ in the comoving frame. As long as photons are trapped inside the fireball, the total energy in the engine frame and entropy in the comoving frame are conserved, from which we can write $V' ({T'})^4 \Gamma\propto R^2 \Gamma^2 ({T'})^4=\rm const$ and $({T'})^3 V'\propto ({T'})^3 R^2\Gamma =\rm const$, respectively. From these two equations, we obtain $T'\approx T_{\rm ini} (R/R_{\rm ini})^{-1}$ and $\Gamma\approx (R/R_{\rm ini})$. 

Let us consider the evolution of a leptonic fireball that consists of electron-positron pairs and photons. As the fireball expands, the temperature becomes less than $m_ec^2$, and after this point, the number density of the pairs decreases exponentially due to the pair-annihilation. The optical depth of the fireball becomes less than 1 when $T'\sim T'_\pm\sim20$ keV \cite{1990ApJ...365L..55S}. At this time, the photospheric radiations are emitted and the fireball loses almost all the energy. The photospheric radius of the leptonic fireball shell is estimated to be $R_\pm\approx R_{\rm ini}(T_{\rm ini}/T'_\pm)\sim5\times10^8 R_{\rm ini,7} (T_{\rm ini}/1\rm ~MeV)$ cm, and the Lorentz factor at this time is $\Gamma_\pm\approx T_{\rm ini}/T'_\pm \sim50 (T_{\rm ini}/1\rm ~MeV)$. The typical photon energy of the photospheric emission is $T_{\rm obs}\sim \Gamma_\pm T'_\pm\sim T_{\rm ini}\sim1~\rm MeV$, which is comparable to the observed gamma-ray energy. However, the spectrum of the photosperic radiation is expected to be similar to the Planck distribution,  which is inconsistent with the broken power-law spectra observed in GRBs. 

Next, we examine a baryonic fireball that contains baryons. The fate of the baryonic fireball is determined by the amount of baryons, which is parameterized by the photon-baryon ratio, $\eta_{\rm fb}=L_{\rm fb}/(\dot{M}_{\rm fb}c^2)$, where $\dot{M}_{\rm fb}$ is the baryonic mass loading rate to the fireball. In this case, the optical depth of the fireball is determined by the electrons accompanied with baryons, whose number density is $n'_e\approx (L_{\rm fb}/\eta_{\rm fb})/(4\pi \Gamma R^2m_pc^3)$. The optical depth is given by $\tau_T\approx \sigma_T n_e' R/\Gamma$, and the photospheric radius at which $\tau_T=1$ is 
\begin{equation}
R_{\rm ph}\approx \frac{L_{\rm fb}\sigma_T}{4\pi m_p c^3 \eta_{\rm fb} \Gamma^2}\sim1\times10^{10}L_{\rm fb,52} \eta_{\rm fb,3}^{-1}\Gamma_3^{-2}\rm~cm.
\end{equation}
On the other hand, if all the initial thermal energy is converted to the kinnetic energy, the terminal Lorentz factor of the fireball is estimated to be $\Gamma_{\rm max}\approx L_{\rm fb}/(\dot{M}_{\rm fb}c^2)=\eta_{\rm fb}$. The coasting radius at which the acceleration stops is 
\begin{equation}
R_{\rm co}\approx \eta_{\rm fb} R_{\rm ini}\sim1\times10^{10}R_{\rm ini,7}\eta_{\rm fb,3}\rm ~ cm.
\end{equation}
If $R_{\rm co} > R_{\rm ph}$, bulk of the thermal energy of the fireball is radiated at the photosphere, resulting in the luminous thermal emission as in the leptonic fireball. Then, the remaining kinetic energy of the fireball is too small to power the non-thermal emission of GRBs.  On the other hand, all the injected energy is converted to the kinetic energy for the case with $R_{\rm co} < R_{\rm ph}$. Then, the kinetic energy of the fireball shell can be converted to the non-thermal particle energy through some dissipation process discussed in the next subsection, which is suitable to explain the luminous and non-thermal photons in GRBs.
 Therefore, the GRB jets are expected to consist of baryons and electrons, and the kinetic energy is dominated by baryons.
The condition of $R_{\rm co} < R_{\rm ph}$ is satisfied when $\eta_{\rm fb} < \eta_*$, where 
\begin{equation}
 \eta_* =\left(\frac{L_{\rm fb}\sigma_T}{4\pi m_pc^3R_{\rm ini}}\right)^{1/4} \simeq1\times10^3 L_{\rm fb,52}^{1/4}R_{\rm ini,7}^{-1/4}.
\end{equation}
The Lorentz factor of the fireball cannot be significantly higher than $\eta_*$. For $\eta_{\rm fb} < \eta_*$, the fireball reaches the terminal Lorentz factor of $\Gamma=\eta_{\rm fb}<\eta_*$. For $\eta_{\rm fb} > \eta_*$, the photopshere appears when $\Gamma\sim\eta_*$, and then, the bulk of the thermal energy is radiated away, making the Lorentz factor of the fireball $\Gamma\sim\eta_*$. From the compactness problem, the GRB jets should have a Lorentz factor of $\Gamma_j > 10^2$. Thus, the Lorentz factor of the GRB jets lies in a range of $10^2 < \Gamma_j \lesssim 10^3$.

\subsection{Internal energy dissipation}

A baryonic fireball converts its initial thermal energy to the kinetic energy. To produce a powerful non-thermal radiation, the kinetic energy should be converted to the non-thermal particle energy via some dissipation processes, such as shocks, turbulence, and magnetic reconnections. A plausible scenario of the conversion mechanism is particle acceleration at the internal shocks formed by the velocity fluctuations among the fireball shells \cite{RM94a,1997ApJ...490...92K}. The shock dissipates kinetic energies of the fireball shells and produce non-thermal electrons through some process, such as diffusive shock acceleration\cite{Dru83a,BE87a}, stochastic acceleration\cite{2012SSRv..173..535P}, or magnetic reconnections \cite{2020PhPl...27h0501G}. Here, we briefly discuss the basic features of the internal shock model.
%This can be done by some dissipation processes, such as internal shocks or magnetic reconnections.
%Accelerated electrons are likely to be responsible for the observed non-thermal gamma-ray spectra. It is widely believed that GRB jets contains protons. If GRB jets consist of only electron-positron pairs, the pair annihilation occurs when its temperature becomes low enough, converting thermal or kinetic energies of the pairs to the radiation energy. Then, most of the jet energies are radiated when the jets become optically thin. The observed gamma-ray spectra are inconsistent with the photons produced by this process. On the other hand, if jets contain hadrons, the kinetic energy of the relativistically moving hadrons are not radiated. Then, 

Suppose that the central engine launches a slow shell of Lorentz factor $\Gamma_s\gg1$ and velocity $\beta_s$ at $t=0$. The engine produces a faster shell of Lorentz factor $\Gamma_r>\Gamma_s$ and velocity $\beta_r$ at $t=t_{\rm eng}$. At the time, the distance between the two shells is $t_{\rm eng}\beta_sc$, and the velocity difference is $(\beta_r-\beta_s)c\approx c/(2\Gamma_s^2)-c/(2\Gamma_r^2)\simeq c/(2\Gamma_s^2)$. The two shells collide each other at $t_d=2t_{\rm eng}\beta_s\Gamma_s^2$, and the distance from the engine is estimated to be
\begin{equation}
R_d=t_d\beta_rc\approx2\Gamma_s^2ct_{\rm eng}\simeq6.0\times10^{13}\rm~cm~\Gamma_{s,2.5}^2t_{\rm eng,-2}.
\end{equation}
The collision forms internal shocks in both shells, which accelerate particles, leading to the gamma-ray emission from the merged shell. The energy and momentum conservations before and after the collision is given by  
\begin{equation}
 M_r\Gamma_r+M_s\Gamma_s = (M_r+M_s+E_m/c^2)\Gamma_m,
\end{equation}
\begin{equation}
 M_r\Gamma_r\beta_r+M_s\Gamma_s\beta_s = (M_r+M_s+E_m/c^2)\Gamma_m\beta_m,
\end{equation}
where $M_r$ and $M_s$ are the mass of the faster and slower shells, respectively, $\Gamma_m$, $\beta_m$, and $E_m$ are the Lorentz factor, velocity, and internal energy of the merged shell, respectively. Rearranging the equations with a condition $\Gamma_r>\Gamma_s\gg1$, we obtain
\begin{equation}
 \Gamma_m=\sqrt{\frac{M_r\Gamma_r+M_s\Gamma_s}{M_r/\Gamma_r+M_s/\Gamma_s}}.
\end{equation}
If the colliding shells have an equal energy, $M_s\Gamma_s=M_r\Gamma_r$, we obtain $\Gamma_m\approx\Gamma_s\sqrt{2/(1+\Gamma_s^2/\Gamma_r^2)}\approx\Gamma_s$. Then, the variability timescale by an observer is given by $t_{\rm var}\approx R_d/(2\Gamma_m^2c)\approx t_{\rm eng}$. Hence, in the internal shock dissipation scenario, the observed variability timescale is roughly equal to the engine variability timescale.% For the cases that the colliding shells have an equal mass, we obtain $\Gamma_m\approx\sqrt{\Gamma_s\Gamma_r}$. Then, the observed variability is  $t_{\rm var}\approx R_d/(2\Gamma_m^2c)\approx(\Gamma_s/\Gamma_r) t_{\rm eng}$.

The dissipated energy can be converted to the non-thermal particle energy, and non-thermal electrons emit gamma-rays via synchrotron radiation. This scenario can reproduce the strongly variable lightcurves of GRBs using the variability of the central engine \cite{1997ApJ...490...92K}. However, it has a few shortcomings. One is the energy conversion efficiency. Detailed calculations revealed that the radiation efficiency is  1\%--10\% \cite{1997ApJ...490...92K,1999ApJ...523L.113K}, which is much lower than the efficiency estimated by observations of prompt and afterglow phases, $\gtrsim10$\% \cite{2007ApJ...655..989Z}.  Another is the minimum value of the low-energy spectral index, $\alpha_1$. The synchrotron process cannot produce photon spectra of $\alpha<2/3$, but some GRBs clearly show much harder spectral indexes based on observations \cite{1998ApJ...506L..23P}.

\section{Neutrinos from Prompt Bursts}\label{sec:prompt}

The prompt phase of GRBs was considered as one of the most promising astrophysical neutrino sources (see Ref. \cite{WB97a} for the pioneering prediction). As seen in the previous section, non-thermal electrons are accelerated in the internal dissipation region of GRB jets. Since the jets should contain protons based on the standard fireball model, protons are likely accelerated together with electrons. Then, the accelerated protons interact with the photons emitted by the electrons, leading to efficient production of high-energy neutrinos via the photomeson production process. In this section, we describe the neutrino production from the prompt phase of GRBs. The current status of neutrino observations and theoretical models are also discussed.

\subsection{Non-thermal proton production}\label{sec:cr-production}

At internal shocks, protons and electrons can be accelerated via diffusive shock acceleration \cite{Dru83a,BE87a}, stochastic acceleration by turbulence \cite{2012SSRv..173..535P}, or magnetic reconnections \cite{2020PhPl...27h0501G}. Here, we do not discuss non-thermal particle acceleration processes in detail. Electrons accelerated in the dissipation region are responsible for emitting the observed gamma-rays through synchrotron emission. Then, the photon spectra emitted by the electrons are described by a broken power-law form as in Equation (\ref{eq:Band}).

In order to calculate the high-energy neutrino emission, we need to know the spectrum of the non-thermal protons, or cosmic-ray (CR) protons, in the emission region. Microscopic processes, such as radiation and acceleration processes, are usually evaluated in the comoving frame. The transport equation of the number spectrum of CR protons, $N_{\varepsilon_p}=dN/d\varepsilon_p$ ($\varepsilon_p$ is the proton energy in the comoving frame), is represented as
\begin{equation}
\frac{\partial  N_{\varepsilon_p}}{\partial t} - \frac{\partial}{\partial \varepsilon_p}\left(\frac{\varepsilon_p N_{\varepsilon_p}}{t_{p,\rm cool}}\right) = \dot{N}_{\varepsilon_p,\rm inj} - \frac{N_{\varepsilon_p}}{t_{p,\rm esc}},
\end{equation}
where $t_{p,\rm cool}$, $\dot{N}_{\varepsilon_p,\rm inj}$, and $t_{i,\rm esc}$, are the cooling time, injection term, and the escape time for protons, respectively, and we use the spatially one-zone approximation. Hereafter, we use $\varepsilon_i$ and $E_i$ for particle energies of species $i$ in the comoving frame and observer frame, respectively.

CR protons are expected to have a power-law spectrum, which is partly confirmed by Particle-In-Cell (PIC) simulations. The acceleration process is usually faster than the dynamical timescales of astrophysical objects. We often assume that CRs of a power-law energy distribution is injected into the emission region:
\begin{equation}
\dot{N}_{\varepsilon_p,\rm inj}= \dot{N}_{p,\rm nor}\left(\frac{\varepsilon_p}{\varepsilon_{p,\rm nor}}\right)^{-s}\label{eq:injection}
\end{equation}
where $s$ is the spectral index, $\varepsilon_{\rm nor}$ is the reference energy of the CR protons, and $\dot{N}_{i,\rm nor}$ is the normalization factor.  The standard diffusive shock acceleration theory predicts $s=2$\cite{BO78a,Bel78a}, and we use it throughout this Chapter. For a typical GRB parameters, the adiabatic expansion is the dominant cooling process, whose timescale is given by the dynamical timescale, $t_{\rm dyn}\simeq R_d/(\Gamma_j c)$. This is independent of particle energy. Balancing the cooling term and injection term with a steady state assumption, the number spectrum is given by $N_{\varepsilon_p}\approx \dot{N}_{\varepsilon_p,\rm inj}t_{\rm dyn}\propto \varepsilon_p^{-s}$. Thus, the power-law index of the number spectrum in the emission region is the same with that for the injection spectrum.

The maximum energy of CR protons is determined by the balance between the acceleration and cooling/escape processes. Phenomenologically,  the acceleration time can be written as $t_{\rm acc}\approx \eta_{\rm acc} r_L/c\ = \eta_{\rm acc} \varepsilon_p/(ceB)$, where $\eta_{\rm acc}$ is the acceleration efficiency parameter, $e$ is the elementary charge, $r_L$ is the Larmor radius, $B\approx \sqrt{2L_{\gamma,\rm iso}\xi_B/(c\Gamma_j^2R_d^2)}\sim8.2\times10^3L_{\gamma,\rm iso,52}^{1/2}\xi_{B,-1}^{1/2}\Gamma_{j,2.5}^{-1}R_{d,14}^{-1}$ G is the magnetic field strength and $\xi_B$ is the energy fraction of the magnetic field compared to the gamma rays.  
Equating the dynamical timescale to the acceleration timescale, the maximum energy of the protons in the observer frame is estimated to be 
\begin{equation}
E_{p,\rm max}\approx\frac{\Gamma_j\varepsilon_{p,\rm max}}{1+z}\approx\frac{eBR_d}{\eta_{\rm acc}(1+z)}\simeq 300\frac{B_4R_{d,14}}{\eta_{\rm acc,0}(1+z)}\rm~EeV.
\end{equation}
Hence, GRBs can accelerate protons to ultrahigh energies with $\eta_{\rm acc}\lesssim10$. Hereafter, we provide $R_d$ as a given parameter, rather than $t_{\rm var}$. 

We provide the normalization, $\dot{N}_{\rm nor}$, in the engine frame. The normalization of the proton spectrum in the engine frame is given by $\mathcal{E}_p=\int \hat{E}_p N_{\hat{E}_p} d\hat{E}_p = \xi_p\mathcal{E}_{\gamma,\rm iso}$, where $\mathcal{E}_p$ is the total proton energy, $\xi_p$ is the baryon loading factor, and $\hat{E}_p$ is the proton energy in the engine frame. For $s=2$, the differential total proton energy in the observer frame is written as
\begin{equation}
 \hat{E}_p^2N_{\hat{E}_p} = \frac{\xi_p 4\pi d_L^2S_{\gamma}}{(1+z)\ln(\hat{E}_{p,\rm max}/\hat{E}_{p,\rm min})},\label{eq:EpNEp}
\end{equation}
where $\hat{E}_{p,\rm min}\approx \Gamma_jm_pc^2$ is the minimum energies of CR protons. With a typical parameter set, the bolometric correction factor is estimated to be $f_{\rm bol}^{-1}=\ln(\hat{E}_{p,\rm max}/\hat{E}_{p,\rm min})\sim15 - 20$. Since $\hat{E}_p^2N_{\hat{E}_p}\propto \varepsilon_p^2 N_{\varepsilon_p}\propto \int \varepsilon_p \dot{N}_{\varepsilon_p,\rm inj}dt$, Equation (\ref{eq:EpNEp}) implicitly provides $\dot{N}_{\rm nor}$.
%The CR proton luminosity in the comoving frame is written as $L'_p = \int \dot{N}_{\varepsilon_p,\rm inj}\varepsilon_p d\varepsilon_p $. The total CR proton energy in the comoving frame, $\mathcal{E}'_p = \mathcal{E}_p/\Gamma_j$, can be estimated to be $\mathcal{E}'_p = \int L'_p  dt'\approx L'_pt'\approx L'_pt_{\rm dur}/\Gamma_j$, where $t'$ is the time in the comoving frame and $t_{\rm dur}$ is the duration of the prompt burst in the engine frame. Using the relations above, we can write $\varepsilon_p^2\dot{N}_{\varepsilon_p,\rm inj}t_{\rm dur} \approx E_p^2 N_{E_p}$.

From the observed intensity and characteristic energy loss length for ultrahigh-energy CRs (UHECRs) of $E\sim10^{19}-10^{20}$ eV, the differential luminosity density of UHECRs is estimated to be $\sim10^{44}\rm~erg~Mpc^{-3}~yr^{-1}$ \cite{2019PhRvD..99f3012M}. The observed GRB rate is $\rho_{\rm GRB}\sim0.1-1\rm~Gpc^{-3}~yr^{-1}$ and the characteristic total gamma-ray energy is $\mathcal{E}_{\rm GRB}\sim10^{53}-10^{54}$ erg  \cite{2010MNRAS.406.1944W,2015ApJ...812...33S}. Then, the luminosity density of gamma-rays is $\rho_{\rm GRB} \mathcal{E}_{\rm GRB}\sim 10^{44}\rm~erg~Mpc^{-3}~yr^{-1}$. Therefore, the energy budget for UHECRs and gamma rays by GRBs are similar. Thus, $\xi_p\sim f_{\rm bol}^{-1}\sim 20$ is demanded in order to explain the observed UHECRs by GRBs \cite{Wax95a,MIN08a}.
%This argument is optimistic in the sense that all the CRs accelerated at the internal dissipation escape from the system without losing energy. As we will see in the next subsection, this is not expected. The most of the accelerated protons lose their energies by adiabatic expansion. Some escape mechanism, such as the neutron conversion is required for CRs to escape from the system. Then, the required baryon loading factor can be $\xi_p\approx f_{\rm bol}^{-1}f_{p\gamma}^{-1}\sim10^3$, where $f_{p\gamma}$ is the pion production efficiency (see the next subsection).

%Since GRB jets contains protons, CR protons are also accelerated at the internal shock or the dissipation region. 
%The spectral index of the CR protons are generally assumed to be $s=2$. This index is theoretically expected in the diffusive shock acceleration process at a non-relativistic shock. In practice, this treatment leads to an efficient neutrino production in all the energy range. The neutrino fluence is high for the energy range where pions are efficiently produced.  

We should note that the proton spectral index can be softer or harder, depending on their acceleration mechanisms and physical environment of dissipation regions. If the CR spectrum is soft ($s>2$), the amount of high-energy neutrinos may be considerably lower than that with $s=2$. Nevertheless, the proton spectrum inside the UHECR source should be $s\lesssim2$, otherwise the required power to explain the observed UHECR energy density is too high. Thus, $s\simeq2$ is suitable to test the GRB-UHECR connection by neutrino observations.

\subsection{High-energy neutrino production}

Non-thermal protons produce various secondary particles, including neutrinos, through photohadronic interactions. The dominant channel of photohadronic interactions in GRBs is pion production. The neutral pions decay to gamma-rays, while charged pions decay to muons and neutrinos. Muons also decay to electrons/positrons and neutrinos. We can write the decay chain as 
\begin{eqnarray}
\rm p+\gamma\rightarrow p (n) +\pi s, \label{eq:decay1}\\
\rm \pi^+\rightarrow\mu^+ + \nu_\mu,\\
\rm \pi^-\rightarrow \mu^- + \overline{\nu}_\mu,\\
\rm \mu^+\rightarrow e^+ + \nu_e +\overline{\nu}_\mu,\\
\rm \mu^-\rightarrow e^-+\overline{\nu}_e + \nu_\mu \label{eq:decay2}
\end{eqnarray}
Neutrinos produced by the decay chain have a flavor ratio of $(\nu_e,~\nu_\mu,~\nu_\tau)\sim(1,~2,~0)$. We expect this flavor ratio for most of astrophysical neutrino sources.

To calculate the neutrino spectrum, we need to know the pion production rate. The cross-section of photomeson production, $\sigma_{p\gamma}$, is given as a function of the photon energy in the proton-rest frame, $\overline{\varepsilon}_{\gamma}=\gamma_p \varepsilon_\gamma (1-\beta_p\mu)$, where $\gamma_p=\varepsilon_p/(m_pc^2)$, $\beta_p$, $\varepsilon_\gamma$,  $\mu=\cos\theta_p$ are the Lorentz factor of protons, the proton velocity, the photon energy, and the angle between the directions of interacting proton and photon in the comoving frame, respectively. The energy loss rate  of a proton by pion production is written as 
\begin{equation}
 t_{p\gamma}^{-1} = c\int d\Omega\int d\varepsilon_\gamma (1-\beta_p\mu)n_\gamma(\varepsilon_\gamma,~\Omega)\sigma_{p\gamma}(\overline{\varepsilon}_{\gamma})\kappa_{p\gamma}(\overline{\varepsilon}_{\gamma}),
\end{equation}
where $n_\gamma(\varepsilon_\gamma,~\Omega)=dN/(d\varepsilon_\gamma dV d\Omega)$ and $\kappa_{p\gamma}(\varepsilon_\gamma)$ is the inelasticity of the photomeson production process (i.e., the fraction of proton energy used to pion production per interaction). This rate is approximately equivalent to the pion production rate.
Photons and non-thermal protons are expected to be isotropic in the comoving frame. Then, the photomeson production rate is represented as \cite{ste68}
\begin{equation}
 t_{p\gamma}^{-1}=\frac{c}{2\gamma_p^2}\int_{\varepsilon_{\rm th}}^{\infty} d\overline{\varepsilon}_\gamma\sigma_{p\gamma}\kappa_{p\gamma}\overline{\varepsilon}_\gamma \int_{\overline{\varepsilon}_\gamma/(2\gamma_p)}^{\infty}\frac{d\varepsilon_\gamma}{\varepsilon_\gamma^2}n_{\varepsilon_\gamma},\label{eq:tpgam}
\end{equation}
where we use $n_\gamma(\varepsilon_\gamma,~\Omega)=n_{\varepsilon_\gamma}/(4\pi)$, $n_{\varepsilon_\gamma}=dN/(d\varepsilon_\gamma dV)$, $\beta_p\sim1$, and convert the integration variable from $\mu$ to $\overline{\varepsilon}_\gamma$.

\subsubsection{Analytic estimates}\label{sec:analytic}

%Ref. \cite{GHA04a} provides neutrino fluences from GRBs with a similar approach, which is used for the first analysis of IceCube GRB search \cite{IceCube10a}. However, their method overestimates the neutrino fluence as shown in Refs. \cite{HBW12a,HLW12a} due to bolometric correction. 
%Here, we follow a formalism provided by Refs. \cite{HLW12a}, in which the bolometric correction is explicitly taken into account.

Ref. \cite{WB97a} provides the first prediction for high-energy neutrinos from GRBs using a simple analytic expression. 
Considering the $\Delta$-resonance process, the crosssection and inelasticity are approximated to be \cite{1979ApJ...228..919S}
\begin{equation}
\sigma_{p\gamma}\kappa_{p\gamma}\approx\sigma_{\rm pk}\kappa_{\rm pk}\Delta\overline{\varepsilon}_{\rm pk}\delta(\overline{\varepsilon}_\gamma-\overline{\varepsilon}_{\rm pk}) \label{eq:resonant}
\end{equation}
where $\sigma_{\rm pk}\sim5\times10^{-28}\rm~cm^{-2}$, $\kappa_{\rm pk}\simeq0.2$, $\overline{\varepsilon}_{\rm pk}\simeq0.3$ GeV are the crosssection, inelasticity, and the photon energy at the resonance peak, $\Delta\overline{\varepsilon}_{\rm pk}\sim0.2$ GeV is the peak width, and $\delta(x)$ is the Dirac delta function. With this approximation, protons of $\varepsilon_p$ interact with photons of $\varepsilon_\gamma\approx(\overline{\varepsilon}_{\rm pk}/\varepsilon_p)m_pc^2$.
By Lorentz transformation of the observed photon spectrum given by Equation (\ref{eq:Band}), we can estimate the differential photon number density to be
\begin{eqnarray}
 n_{\varepsilon_\gamma}\approx\frac{\chi_\alpha u_\gamma}{\varepsilon_{\gamma,\rm br}^2}\left\{
\begin{array}{lc}
 \left(\frac{\varepsilon_\gamma}{\varepsilon_{\gamma,\rm br}}\right)^{-\alpha_1}~~ & ~~(\varepsilon_\gamma<\varepsilon_{\gamma,\rm br}) \\
 \left(\frac{\varepsilon_\gamma}{\varepsilon_{\gamma,\rm br}}\right)^{-\alpha_2}~~ & ~~(\varepsilon_\gamma>\varepsilon_{\gamma,\rm br}) 
\end{array}
\right.\label{eq:dndegam}
\end{eqnarray}
where $\chi_\alpha=(2-\alpha_1)(\alpha_2-2)/(\alpha_2-\alpha_1)$ is the bolometric correction of the photon spectrum\footnote{Here, we assume that the broken power-law photon spectrum continues to the lower and higher energy bands, and we define $L_{\gamma,\rm iso}$ as the total photon luminosity, $L_{\gamma,\rm iso} = 4\pi \Gamma_j^2R^2 c\int d\varepsilon_\gamma \varepsilon_\gamma n_{\varepsilon_\gamma} $. This is different from the gamma-ray luminosity in the observed band. In reality, the observation has a limited band width. We need to correct the gamma-ray luminosity in the observed band width to an intrinsic gamma-ray luminosity by extrapolating the broken power-law spectrum to estimate the baryon loading factor. This correction may affect the resulting baryon loading factor by a factor of a few to several \cite{KMM17b}.  }, $u_\gamma=\int d\varepsilon_\gamma \varepsilon_\gamma n_{\varepsilon_\gamma}\approx L_{\gamma,\rm iso}/(4\pi R_d^2\Gamma_j^2c)$ is the photon energy density, $\varepsilon_{\gamma,\rm br}=(1+z)E_{\gamma,\rm br}/\Gamma_j$ is the photon break energy in the comoving frame, and we assume $\alpha_1<2$ and $\alpha_2>2$. Substituting Equations (\ref{eq:resonant}) and (\ref{eq:dndegam}) into Equation (\ref{eq:tpgam}), we obtain
\begin{equation}
 t_{p\gamma}^{-1}\approx \frac{2\chi_\alpha u_\gamma c \sigma_{\rm pk}\kappa_{\rm pk}\Delta\overline{\varepsilon}_{\rm pk}}{\overline{\varepsilon}_{\rm pk}\varepsilon_{\gamma,\rm br}(1+\alpha_1)}\left\{
\begin{array}{lc}
   \chi_1 \left(\frac{\varepsilon_p}{\varepsilon_{p,\rm br}}\right)^{\alpha_2-1}~~  & ~~ (\varepsilon_p<\varepsilon_{p,\rm br}) \\
\left(\frac{\varepsilon_p}{\varepsilon_{p,\rm br}}\right)^{\alpha_1-1}+\chi_2~~  & ~~ (\varepsilon_p>\varepsilon_{p,\rm br}) \\
\end{array}
\right.,
\end{equation}
where $\varepsilon_{p,\rm br}=\overline{\varepsilon}_{\rm pk}m_pc^2/(2\varepsilon_{\gamma,\rm br})$, $\chi_1=(1+\alpha_1)/(1+\alpha_2)$ and $\chi_2=(\varepsilon_p/\varepsilon_{p,\rm br})^{-2}(\alpha_1-\alpha_2)/(1+\alpha_2)$.
%, we omit these factors in the following equations\footnote{The line in Figure \ref{fig:comparison} appropriately take these factors into account.}. 

The CR protons lose energies by adiabatic expansion in a dynamical timescale, $t_{\rm dyn}\approx R_d/(\Gamma_jc)$. The pion production efficiency, i.e., the fraction of CR protons producing pions through the photomeson production, is given by $f_{p\gamma}={\rm min}(1,t_{p\gamma}^{-1}/t_{\rm dyn}^{-1})$. Converting other quantities into the observer frame, we obtain
\begin{equation}
 f_{p\gamma}\approx\frac{\chi_\alpha\sigma_{\rm pk}\kappa_{\rm pk}\Delta\overline{\varepsilon}_{\rm pk}L_\gamma}{2\pi c\overline{\varepsilon}_{\rm pk}\Gamma_j^2R_dE_{\gamma,\rm br}(1+\alpha_1)(1+z)}\left\{
\begin{array}{lc}
  \chi_1\left(\frac{E_p}{E_{p,\rm br}}\right)^{\alpha_2-1}~~  & ~~ (E_p<E_{p,\rm br}) \\
  \left(\frac{E_p}{E_{p,\rm br}}\right)^{\alpha_1-1}+\chi_2~~  & ~~ (E_p>E_{p,\rm br}) \\
\end{array}
\right.,
\end{equation}
\begin{equation}
E_{p,\rm br}=\frac{\Gamma_j^2\overline{\varepsilon}_{\rm pk}m_pc^2}{2E_{\gamma,\rm br}(1+z)^2}
\end{equation}
Using typical values for observed GRBs, we can write
\begin{eqnarray}
& & f_{p\gamma}\simeq0.2 \frac{\chi_\alpha L_{\gamma,52}}{\Gamma_{j,2.5}^2R_{d,14}}\left(\frac{E_{\gamma,\rm br}}{300\rm~keV}\right)^{-1}\left(\frac{2}{1+\alpha_1}\right)\left(\frac{2}{1+z}\right)\left\{
\begin{array}{lc}
  \chi_1\left(\frac{E_p}{E_{p,\rm br}}\right)^{\alpha_2-1}~~  & ~~ (E_p<E_{p,\rm br}) \\
  \left(\frac{E_p}{E_{p,\rm br}}\right)^{\alpha_1-1}+\chi_2~~  & ~~ (E_p>E_{p,\rm br}) \\
\end{array}
\right.,\nonumber \\
& &E_{p,\rm br}\simeq1.2\times10^{16}\Gamma_{j,2.5}^2\left(\frac{E_{\gamma,\rm br}}{300\rm~keV}\right)^{-1}\left(\frac{1+z}{2}\right)^{-2}\rm~eV.~~~\nonumber
\end{eqnarray}
Therefore, a considerable fraction of accelerated CR proton energies can be spent to produce pions. Since $\chi_1\sim1$ and $\chi_2\ll1$ unless $\varepsilon_p\sim\varepsilon_{p,\rm br}$, these factors do not affect the results significantly. In contrast, $\chi_\alpha=1/6\simeq0.17$ for $\alpha_1=1$ and $\alpha_2=2.2$, and hence, we cannot ignore $\chi_\alpha$ for the estimate.

Roughly half of the $p\gamma$ interactions produce charged pions, whose energy is $\varepsilon_\pi\approx0.2\varepsilon_p$. Thus, the differential pion production rate can be written as $\dot{N}_{\varepsilon_\pi,\rm inj}\approx (1/2)f_{p\gamma} \dot{N}_{\varepsilon_p,\rm inj}$. The higher energy pions can lose their energies by either synchrotron radiation or dynamical expansion before they decay to neutrinos. The transport equation for the pions in a steady state can be written as 
\begin{equation}
 -\frac{d}{d\varepsilon_\pi}\left(\frac{\varepsilon_\pi N_{\varepsilon_\pi}}{t_{\pi,\rm cl}}\right) =\dot{N}_{\varepsilon_\pi,\rm inj} -\frac{N_{\varepsilon_\pi}}{t_{\rm dec}},\label{eq:transpi}
\end{equation}
where  $t_{\pi,\rm cl}$ is the pion cooling time, $t_{\pi,\rm dec}=\varepsilon_\pi \tau_\pi/(m_\pi c^2)$ is the decay time of charged pions, and $\tau_\pi\simeq2.6\times10^{-8}$ sec is the decay time of charged poins at rest. In typical prompt phases of GRBs, synchrotron cooling is the most efficient, whose cooling timescale is $t_{\pi,\rm syn}=6\pi m_\pi^4c^3/(m_e^2\sigma_TB^2\varepsilon_\pi)$. From $t_{\pi,\rm syn}=t_{\rm dec}$, we can write the pion cooling energy in the observer frame as
\begin{equation}
 E_{\pi,\rm cl}=\frac{\Gamma_j\varepsilon_{\pi,\rm cl}}{1+z}=\frac{\Gamma_j}{1+z}\sqrt{\frac{6\pi m_\pi^5c^5}{m_e^2\sigma_T\tau_\pi B^2}}\simeq3.4\times10^{18}\Gamma_{j,2.5}B_4^{-1}(1+z)^{-1}\rm~eV.
\end{equation}
For $\varepsilon_\pi<\varepsilon_{\pi,\rm cl}$, the cooling is ineffective, and thus, all the injected pions decay, leading to the decay term $N_{\varepsilon_\pi}/t_{\rm dec}\approx\dot{N}_{\varepsilon_\pi,\rm inj}$. On the other hand, the cooling term balances the injection term for $\varepsilon_\pi>\varepsilon_{\pi,\rm cl}$. This leads to $N_{\varepsilon_\pi}\sim \dot{N}_{\varepsilon_\pi,\rm inj}t_{\pi,\rm cl}$, where we assume a power-law injection term. Then, the decay rate is modified to $N_{\varepsilon_\pi}/t_{\pi,\rm dec}\approx(t_{\pi,\rm cl}/t_{\pi,\rm dec})\dot{N}_{\varepsilon_\pi,\rm inj}=(\varepsilon_{\pi,\rm cl}/\varepsilon_\pi)^2\dot{N}_{\varepsilon_\pi,\rm inj}$, where we assume that the synchrotron cooing is dominant in the last equation.
 The pions decay to four particles ($\pi^\pm \rightarrow 3\nu + e^\pm$) that equally share the parent pion energy, leading to $\varepsilon_{\nu_\mu}\approx \varepsilon_\pi/4 \approx 0.05 \varepsilon_p$. The muon neutrino production rate is given by $\dot{N}_{\varepsilon_{\nu_\mu},\rm inj}\approx (1/4)N_{\varepsilon_\pi}/t_{\pi,\rm dec}$. The differential total muon neutrino energy, $\hat{E}_\nu^2N_{\hat{E}_\nu}\propto\int \varepsilon_{\nu_\mu}^2\dot{N}_{\varepsilon_{\nu_\mu},\rm inj}dt$, is then approximated to be
\begin{eqnarray}
 \hat{E}_\nu^2N_{\hat{E}_\nu}\approx\frac{1}{8}f_{p\gamma}f_{\pi,\rm sup}{\hat{E}_p}^2N_{\hat{E}_p},\label{eq:dNdEpi}\\
 f_{\pi,\rm sup}\approx 1-\exp\left(-\frac{t_{\pi,\rm cl}}{t_{\pi,\rm dec}}\right), \label{eq:fpicl}
%\left\{
% \begin{array}{lc}
%1  & (E_\pi < E_{\pi,\rm cl}) \\
%\left(\frac{E_\pi}{E_\pi,\rm cl}\right)^{-2}  & (E_\pi > E_{\pi,\rm cl}) \\
% \end{array}
%\right.,
\end{eqnarray}
where we introduce the pion cooling suppression facotore $f_{\pi,\rm sup}$ and use the quantities in the engine frame in order to normalize the neutrino emission energy. We can approximate $f_{\pi,\rm sup}$  to the two asymptotes, $f_{\pi,\rm sup}\approx1$ for $ E_\pi < E_{\pi,\rm cl}$ and $f_{\pi,\rm sup}\approx(E_\pi/E_{\pi,\rm cl})^2$ for $ E_\pi > E_{\pi,\rm cl}$. We can write down the muon neutrino spectrum fully analytically with this prescription.

\begin{figure}[tb]
  \begin{minipage}[c]{0.48\linewidth}
 \begin{center}
  \includegraphics[width=\linewidth]{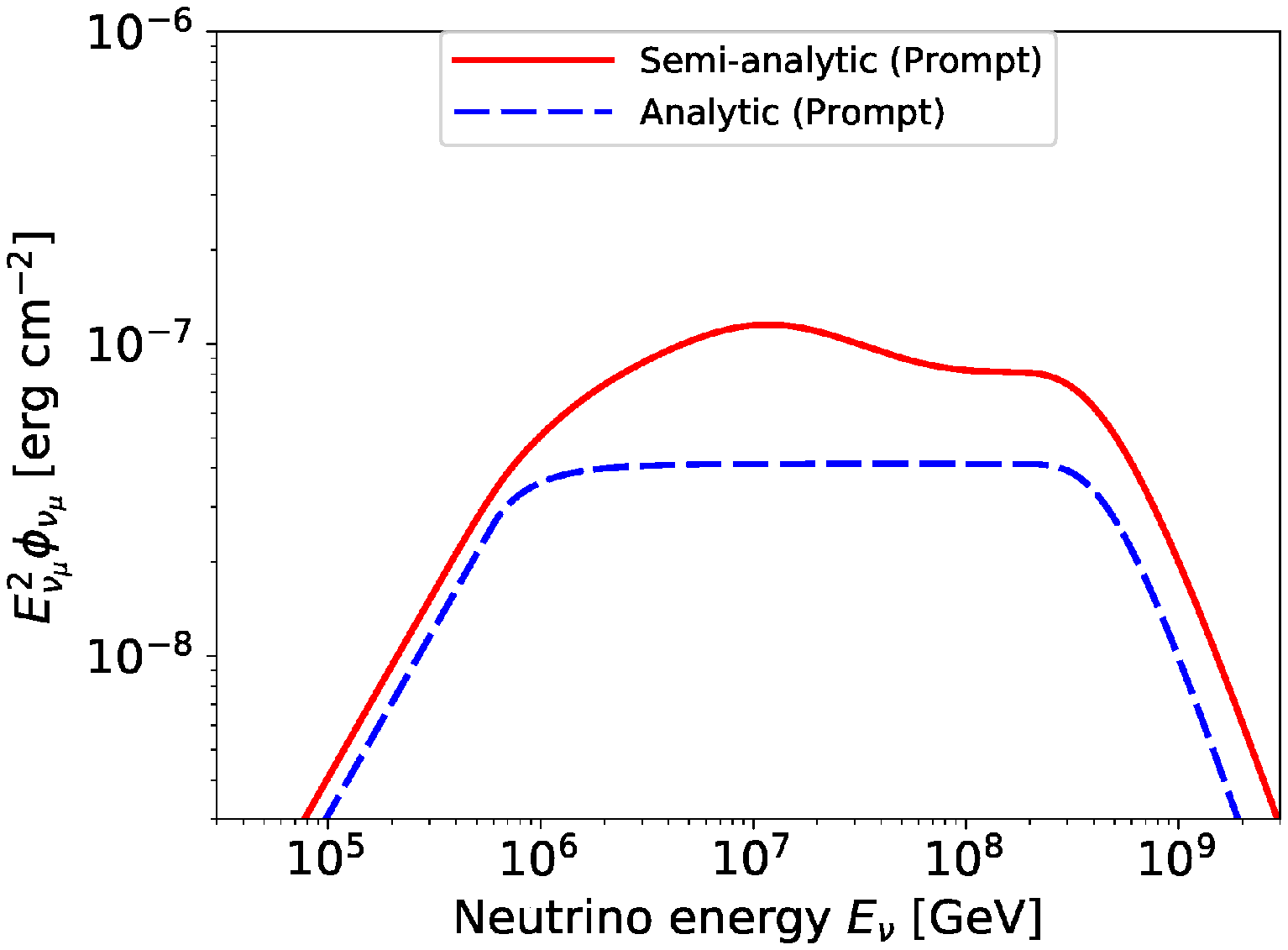}
 \end{center}   
  \end{minipage}
  \begin{minipage}[c]{0.48\linewidth}
 \begin{center}
  \includegraphics[width=\linewidth]{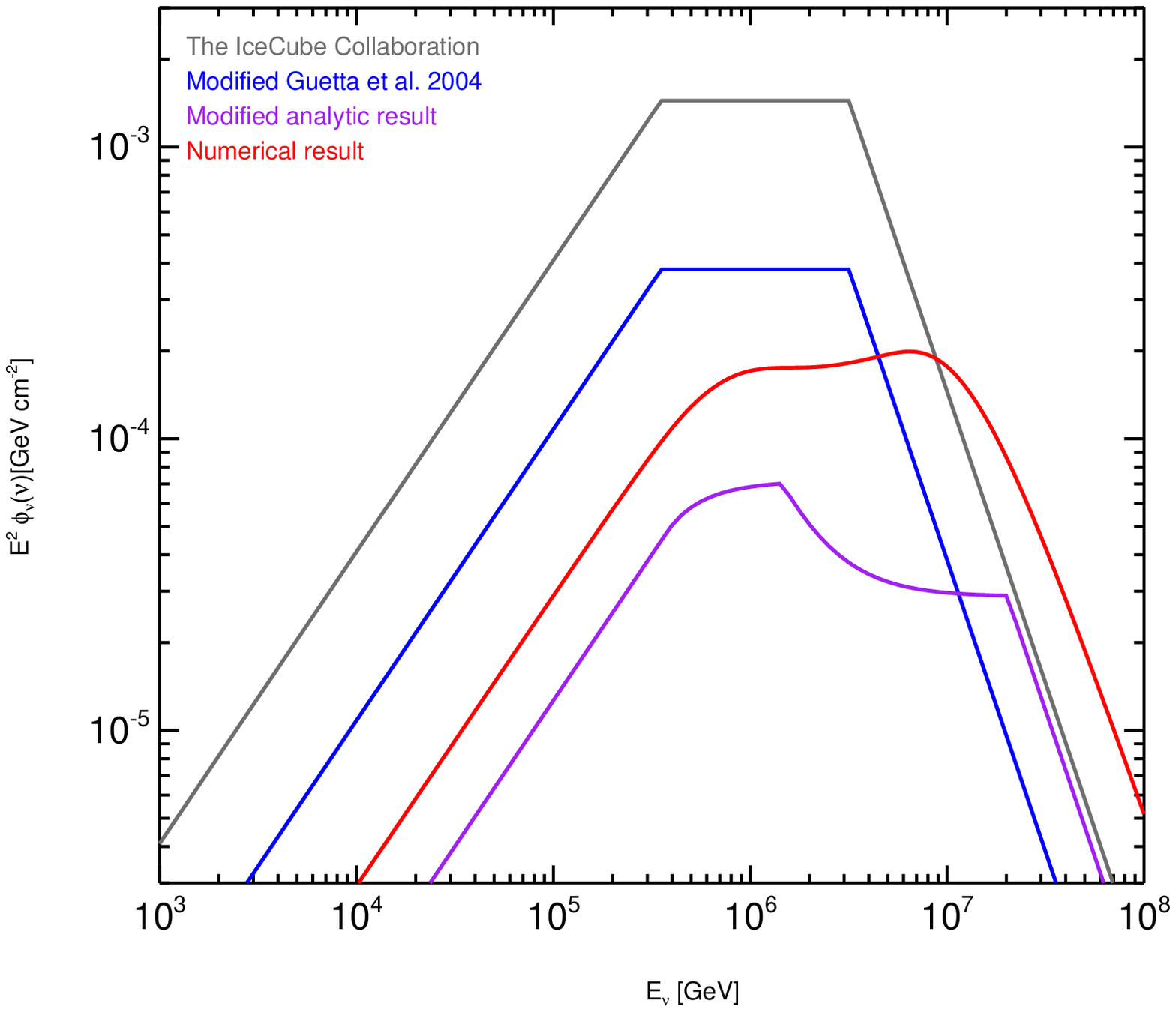}
 \end{center}      
  \end{minipage}
  \caption{Left: Differential muon neutrino fluences from a typical GRB with the analytic estimate in Section \ref{sec:analytic} (dashed) and the semi-analytic approach in Section \ref{sec:semi-analytic} (solid). Parameters are $L_{\gamma,\rm iso}=10^{52}\rm~erg~s^{-1}$, $\Gamma_j=10^{2.5}$, $\mathcal{E}_{\gamma,\rm iso}=10^{53}$ erg, $E_{\gamma,\rm br}=300$ keV, $R_d=10^{14}$ cm, $\alpha_1$=1.0, $\alpha_2$=2.2, $\xi_B=0.1$, $\xi_p=10$, $\eta_{\rm acc}=1$, and $z=1.0$ (corresponding to $d_L\simeq6.6$ Gpc).  
Right: Same as the left panel, but for four different methods (method by Ref. \cite{IceCube10a}), Ref. \cite{GHA04a}, semi-analytic method in Ref. \cite{HLW12a}, and numerical method in Ref. \cite{HLW12a}. The right panel is reproduced from Ref. \cite{HLW12a} with permission by AAS.
 }\label{fig:comparison}
\end{figure}

The left pnael of Figure \ref{fig:comparison} shows the muon neutrino energy fluence from a typical GRB, 
\begin{equation}
 E_{\nu_\mu}^2\phi_{\nu_\mu} = \frac{(1+z){\hat{E}_\nu}^2 N_{\hat{E}_\nu}}{4\pi d_L^2} = 
 \frac18 f_{p\gamma}f_{\pi,\rm sup}\frac{\xi_p S_\gamma}{\ln(E_{p,\rm max}/E_{p,\rm min})}.\label{eq:muonSED}
\end{equation}
With this simple treatment, the neutrino fluence is  a power-law shape with two break points. At the lowest energy branch, CR protons interact with photons of $E_\gamma>E_{\gamma,\rm br}$, which leads to $E_\nu^2 \phi_\nu\propto E_\nu^{2-s+\alpha_2-1}$. The lower-energy break at $E_{\nu,\rm br}=0.05E_{p,\rm br}$ is caused by the break of the target photon spectrum. Protons above the energy can interact with the lower energy photons that have a harder spectrum, making the neutrino spectrum softer: $E_\nu^2 \phi_\nu\propto E_\nu^{2-s+\alpha_1-1}$. The higher-energy break at $E_{\nu,\rm cl}=0.25 E_{\pi,\rm cl}$ is due to the pion cooling, above which the pion synchrotron cooling suppresses the neutrino production, which results in $E_\nu^2  \phi_\nu\propto E_\nu^{2-s+\alpha_1-3}$. This type of simple treatments were widely used for the prediction for neutrino fluence from individual GRBs~\cite{GHA04a} and early phase of the GRB neutrino analysis~\cite{IceCube10a,Abbasi:2011qc} (see Refs. \cite{HBW12a,2012PhRvD..85b7301L,HLW12a} for the issues regarding the initial GRB analyses). 

\subsubsection{Semi-analytic approach}\label{sec:semi-analytic}

The analytic treatment enables us to predict the neutrino fluence with a simple calculation, but numerical treatments are required in order to predict the neutrino spectra more accurately. Here, we introduce a semi-analytic method used in Refs. \cite{KMM17b,KMB18a}, which gives a more accurate outcome and still straightforward to implement. This method adds a few physical processes that are ignored in the analytic treatment.

One is the cross-section of the photomeson production. We should include not only the delta-resonance channel, but also the multi-pion channel. This channel is relevant for interactions with $\overline{\varepsilon}_\gamma > \overline{\varepsilon}_{\rm pk}$, and for a hard target photons, it can dominate over the resonant channel. Because of the hard target photon spectra in GRBs, the neutrino production efficiency is considerably modified up to a factor of $\sim3$. We should numerically integrate over Equation (\ref{eq:tpgam}) with a full cross-section of photomeson production (see e.g., Ref. \cite{2000CoPhC.124..290M}). 

The muon cooling also affects the neutrino spectra in a way similar to the pion cooling. The muon cooling suppresses the production of electron and anti-muon neutrinos. For the muon neutrinos produced by pion decay, the spectrum is given by Equations (\ref{eq:muonSED}). The anti-muon and electron neutrinos produced by muon decay is given  by
\begin{equation}
 E_{\nu_e}^2\phi_{\nu_e}\approx E_{\overline{\nu}_\mu}^2\phi_{\overline{\nu}_\mu} \approx \frac18 f_{p\gamma}f_{\pi,\rm sup}f_{\mu,\rm sup} \frac{\xi_p S_\gamma}{\ln(E_{p,\rm max}/E_{p,\rm min})}
\end{equation}
where $E_{\nu_e}\approx E_{\overline{\nu}_\mu}\approx0.05E_p$ are the neutrino energies and $f_{\mu,\rm sup}\approx 1-\exp(-t_{\mu,\rm cl}/t_{\mu,\rm dec})$ is the suppression factor by the muon cooling. In addition to the synchrotron cooling, the adiabatic loss is taken into account by using $t_{i,\rm cl}^{-1} = t_{i,\rm syn}^{-1}+t_{\rm dyn}^{-1}$. The mass of a muon is similar to that of a pion, but muon's lifetime at the particle rest frame is two orders of magnitude shorter than that for pions, and hence, the critical energy of muon coolins suppression is approximately $\varepsilon_{\mu,\rm cl}\sim0.1\varepsilon_{\pi,\rm cl}$. Thus, the neutrino spectrum contains an energy range in which the pion cooling is negligible but muon cooling is effective. In the energy range, only the muon neutrinos are efficiently produced, and then, the flavor ratio in the energy range is $(\nu_e,~\nu_\mu,~\nu_\tau)\sim(0,~1,~0)$ at the source.

Another process that affects the neutrino spectra is neutrino oscillation, or flavor mixing, which changes the neutrino flavor ratio during the propagation from the source to the Earth. For astrophysical neutrinos, we can use the long distance approximation for the oscillation factor, $\sin^2(\delta m_{ij}L/(4E_\nu))\sim 0.5$. Then, if we make approximation for $\theta_{12}\sim\pi/6$, $\theta_{23}\sim\pi/4$, and $\theta_{13}\sim0$\footnote{The best-fit parameters in the current experiment is $\theta_{12}\simeq0.58$, $\theta_{23}\simeq0.86$, and $\theta_{13}\simeq0.15$ \cite{2020JHEP...09..178E}. This slightly affects the resulting flavor ratio, but the difference from Equation (\ref{eq:mixing}) is at most a factor of 1.5.}, we can write the relation between the neutrino fluence at the source and neutrino fluence on Earth as \cite{bec08}
\begin{equation}
 \left(
\begin{array}{c}
      \phi_{\nu_e}^{\rm Earth} \\
      \phi_{\nu_\mu}^{\rm Earth} \\
      \phi_{\nu_\tau}^{\rm Earth} \\
\end{array}
\right) 
= \frac{1}{18}\cdot
 \left(
\begin{array}{ccc}
      10 & 4 & 4 \\
      4 & 7 & 7 \\
      4 & 7 & 7 \\
\end{array}
\right) \cdot
 \left(
\begin{array}{c}
      \phi_{\nu_e}^{\rm Source} \\
      \phi_{\nu_\mu}^{\rm Source} \\
      \phi_{\nu_\tau}^{\rm Source} \\
\end{array}
\right) .\label{eq:mixing}
\end{equation}
The flavor ratio of neutrinos produced by pion decay is $(\nu_e,~\nu_\mu,~\nu_\tau)=(1,~2,~0)$ at the source without muon cooling. This ratio is changed to $(\nu_e,~\nu_\mu,~\nu_\tau)\simeq(1,~1,~1)$ on Earth.
 With an efficient muon cooling, the flavor ratio is $(\nu_e,~\nu_\mu,~\nu_\tau)=(0,~1,~0)$ at the source, which results in $(\nu_e,~\nu_\mu,~\nu_\tau)=(1,~1.75,~1.75)$ on Earth.

 The neutrino fluence calculated with the semi-analytic treatment is also shown in the left panel of Figure \ref{fig:comparison}. We can see that the resulting fluence is a factor of 2-3 higher than that by the analytic estimate at the peak energy. This is mainly due to the increase of crossection by the multi-pion channel. The increase of the crosssection results in a higher pion production rate at $E_\nu\gtrsim10^6$ GeV, and indeed, the muon neutrino spectrum at the source indicates a hard spectrum as long as $E_\nu < E_{\nu,\rm cl}$. The muon cooling suppresses the neutrino production for $E_\nu \gtrsim 0.1 E_{\nu,\rm cl}\sim 10^7$ GeV, compensating the high pion production rate at the energy. This makes a fairly flat spectrum for $E_\nu \sim3\times10^6-3\times10^8$ GeV, and the spectral shape of the numerical result is similar to that by the analytic estimate.

We should note that the difference between the analytic and numerical results depends on parameters, and thus, a numerical treatment is always favorable in order to constrain some parameters using the observational data. The neutrino spectra from GRBs with further details, including the distribution of the secondary particles and contributions by other mesons, are shown in Refs. \cite{MN06b,Baerwald:2010fk}. These results are similar to that by the semi-analytic method, but difference can be as large as a factor of 2--3. The right panel of Figure \ref{fig:comparison} shows a detailed comparison between analytic methods and a fully numerical method~\cite{HLW12a}, from which we see that the neutrino fluence can be 1-2 orders of magnitude different depending on the treatment. See also Ref. \cite{HBW12a} for detailed comparison between the analytic and numerical methods.

For detection of astrophysical neutrinos, the main background is the atmospheric neutrinos. They rapidly decrease with the neutrino energy, and at the energy range of GRB neutrino ($E_\nu\sim1-100$ PeV), the atmospheric background is negligibly small. Also, from electromagnetic observations, we can obtain the position and the time frame of the GRB events, which is useful to reduce the atmospheric background.  Hence, GRBs are suitable target to search for the astrophysical neutrinos. The GRB neutrinos by the analytic method used to predict detection of more than 10 neutrinos with a cubic-kilometer detector if 500 GRBs were stacked \cite{Ahrens:2002dv}, but currently no GRB neutrinos are reported as discussed in Section \ref{sec:IceCubeGRB}.

\subsection{Prediction for cosmic high-energy neutrino background}

GRBs are expected to emit high-energy neutrinos of PeV energies. Integrating over the neutrino emissions from all the GRBs in the Universe, we can estimate the contribution to the cosmic high-energy neutrino background by GRBs. As discussed in Section \ref{sec:cr-production}, the total cosmic-ray energy is normalized using the baryon loading factor as $\mathcal{E}_{p,\rm iso}=\xi_p \mathcal{E}_{\gamma,\rm iso}$. For a typical value of $\alpha_1\sim1$, the pion production efficiency is independent of the neutrino energy. The differential total neutrino energy for $E_{\nu,\rm br} < E_\nu < E_{\nu,\rm cl}$ for a typical GRB is then estimated to be
\begin{eqnarray}
& &E_\nu^2 N_{E_\nu}\approx \frac38 f_{\rm bol}f_{p\gamma}\xi_p\mathcal{E}_{\gamma,\rm iso}\\
& &\sim 6\times10^{51}\left(\frac{f_{p\gamma}}{0.05}\right)\left(\frac{\xi_p}{20}\right)\left(\frac{f_{\rm bol}}{0.05}\right)\left(\frac{\mathcal{E}_{\gamma,\rm iso}}{3\times10^{53}\rm~erg}\right)\rm~erg.\nonumber
\end{eqnarray}
The differential energy budget of high-energy neutrinos is $\rho_{\rm GRB} E_\nu^2N_{E_\nu}\simeq6\times10^{42}\rm~erg~Mpc^{-3}~yr^{-1}$. GRBs occur across the  cosmic time, and the differential neutrino energy density produced by GRBs is approximately given by $U_{E_\nu}\approx f_z\rho_{\rm GRB}E_\nu N_{E_\nu}/H_0$, where $H_0\sim70\rm~km~s^{-1}~Mpc^{-1}$ and $f_z\sim3$ is the correction factor due to redshift evolution \cite{WB99a}. Then, the all-flavor neutrino intensity of the diffuse component by GRBs is given by 
\begin{eqnarray}
& & E_\nu^2\Phi_\nu\approx\frac{c}{4\pi}E_\nu U_{E_\nu}\approx\frac{c}{4\pi H_0}\frac38f_zf_{\rm bol}f_{p\gamma}\xi_p\mathcal{E}_{\gamma,\rm iso}\rho_{\rm GRB} \\
& &\sim1\times10^{-8}{\rm~GeV~s^{-1}~cm^{-2}~sr^{-1}}\left(\frac{f_{p\gamma}}{0.05}\right)\left(\frac{\xi_pf_{\rm bol}}{1}\right)\left(\frac{f_z}{3}\right)\left(\frac{\rho_{\rm GRB}\mathcal{E}_{\gamma,\rm iso}}{3\times10^{44}\rm~erg~Mpc^{-3}~yr^{-1}}\right).\nonumber
\end{eqnarray}
This is detectable by a cubic kilometer detector, and thus, GRBs were considered as a promising source candidate of high-energy cosmic neutrinos.

\subsection{IceCube Constraints}\label{sec:IceCubeGRB}

\begin{figure}[tb]
  \begin{minipage}[c]{0.54\linewidth}
 \begin{center}
  \includegraphics[width=\linewidth]{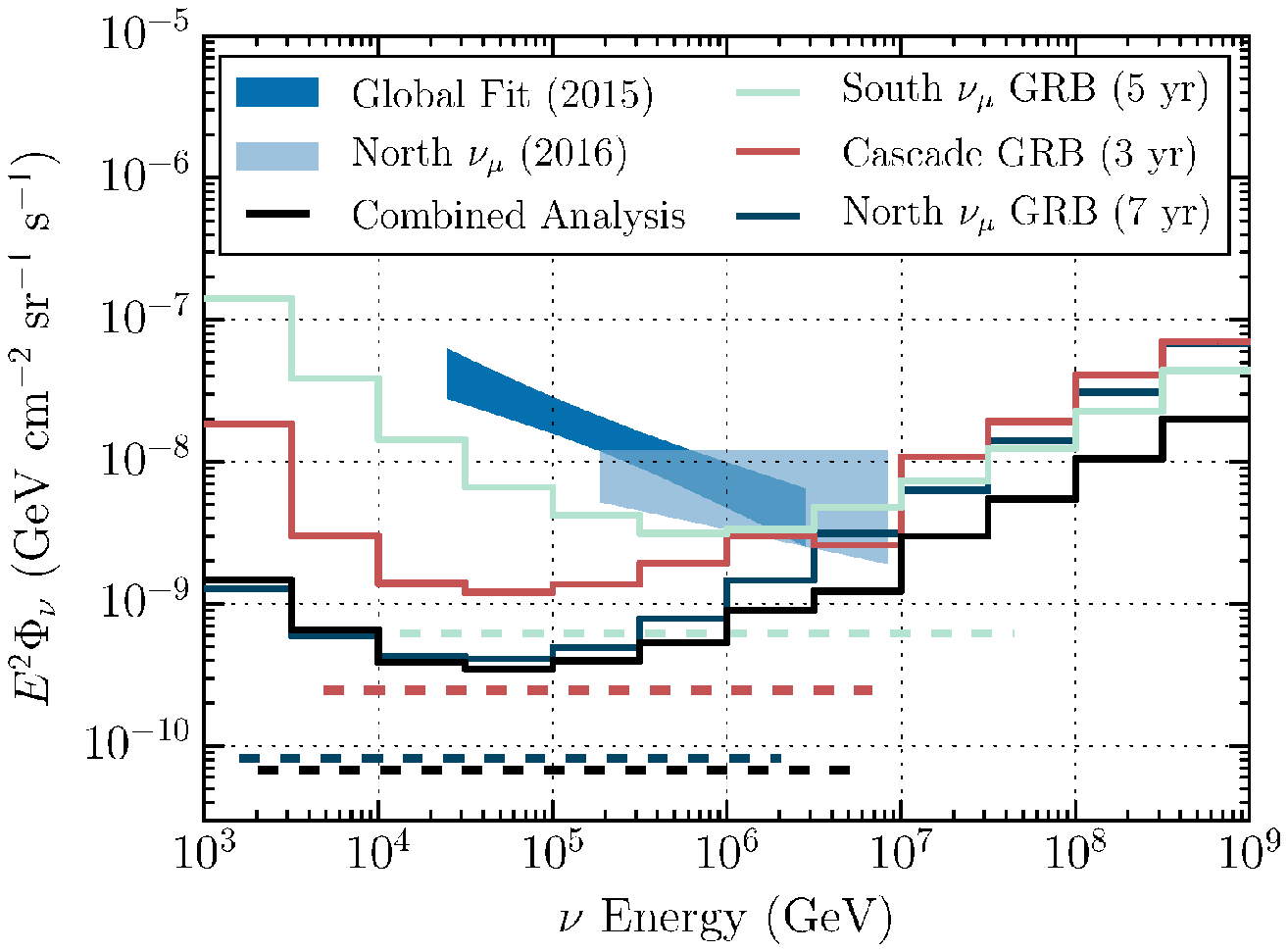}
 \end{center}   
  \end{minipage}
  \begin{minipage}[c]{0.43\linewidth}
 \begin{center}
  \includegraphics[width=\linewidth]{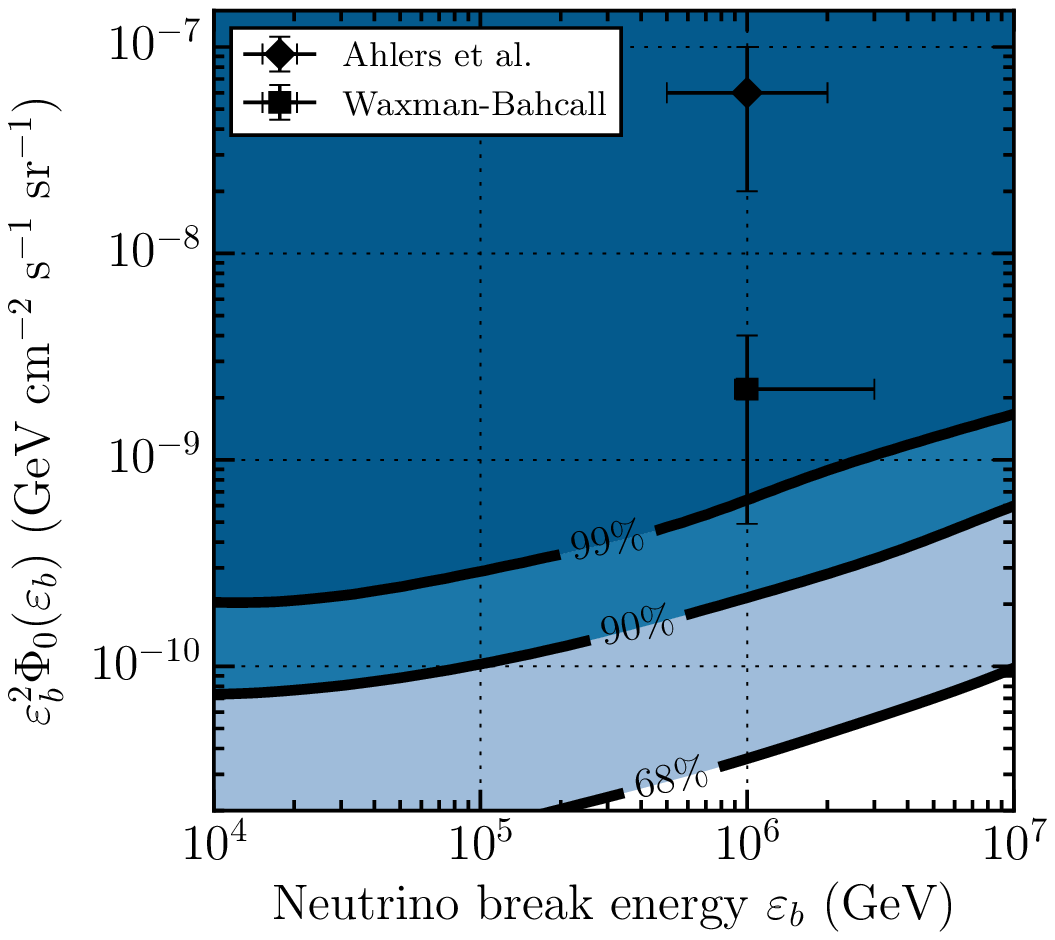}
 \end{center}      
  \end{minipage}
  \caption{Left: Constraints on the GRB contribution to the diffuse neutrino intensity. The solid lines are the upper limit by the energy dependent analyses for track events in southern sky, cascade events in all sky, tracks in the northern sky, and combined analysis shown in black. The dashed lines are limit by the energy integrated analyses. The shaded regions indicate the diffuse neutrino intensities by the analyses shown in the caption.
 Right: Constraint on parameters of the double-broken power-law model discussed in Section \ref{sec:analytic}. The horizontal axis is the lower break energy, $E_{\nu,\rm br}$, and the vertical axis is the normalization of the diffuse neutrino intensity, $E_\nu^2 \phi_\nu$ above the break energy. The two points show the parameter space of GRB internal shock models as the origin of UHECRs \cite{WB97a,2011APh....35...87A}.
 Both panels are reproduced from Ref. \cite{IceCube17b} with permission by AAS.
 }\label{fig:constraints}
\end{figure}

IceCube detector was built at the South Pole in order to detect astrophysical neutrinos. The IceCube Collaboration performed the dedicated analyses to search for neutrinos associated with GRBs. However, their results are consistent with no associated neutrino so far \cite{Abbasi:2011qc,IceCube12a,IceCube15a,IceCube16b,IceCube17b}. In the latest analysis\cite{IceCube17b}, they included more than 1000 electromagnetically detected GRBs and 5 years of the IceCube neutrino data, and put a stringent constraint on the diffuse neutrino flux from GRBs. The neutrino flux from GRBs should be less than 1\% of the cosmic high-energy neutrino background detected by IceCube \cite{ice15a,Aartsen:2020aqd} (see the left panel of Fig. \ref{fig:constraints}), which is surprisingly low. The neutrino fluence from the internal shock scenario primarily depends on the Lorentz factor and the baryon loading factor. With a standard Lorentz factor of $\Gamma_j\sim300$, they obtain $\xi_p<3$ for the 90\% confidence level. $\Gamma_j\gtrsim600$ is necessary to allow the required value for GRBs to be the source of UHECRs, $\xi_p\sim30$. The right panel of Figure \ref{fig:constraints} indicates that IceCube non-detection ruled out classical GRB-UHECR models given by Refs. \cite{WB97a,2011APh....35...87A}.

Neutrino fluences can be different if we consider a different scenario of internal dissipation. Ref. \cite{2013PhRvL.110l1101Z} discusses the neutrino fluence from three different scenarios: internal shock, dissipative photosphere \cite{2005ApJ...628..847R} (see Ref. \cite{Mur08a} for discussion on neutrino emission), and the Internal Collsion-induced Magnetic Reconnection and Turbulence (ICMART) \cite{2011ApJ...726...90Z}. Each scenario has a different dissipation radius. The dissipative photosphere scenaio has the smallest dissipation radius, and thus, the IceCube can put the harshest constraint with no allowed region of $\xi_p\sim30$ for $\Gamma_j<10^3$. On the other hand, ICMART scenario has larger dissipation radii, which can reduce the required Lorentz factor down to $\Gamma_j\sim250$ in order to accommodate $\xi_p\sim30$.

ANTARES, a neutrino detector in Mediterranean Sea, is also searching for neutrinos from GRBs. IceCube suffers from the strong atmospheric background for GRBs in the southern hemisphere, and thus, ANTARES has a better sensitivity to them. ANTARES Collaboration discovered no associated neutrino\cite{Adrian-Martinez:2013dsk,Albert:2020lvs}. Although their constraints are weaker than that by IceCube, the GRB contribution to the cosmic high-energy neutrino background is independently constrained to be less than 10\%.

\subsection{Beyond one-zone approximation}

In reality, the central engine of GRBs launches a number of relativistic shells, which collides at various radii. Then, the internal dissipation also occurs at multiple regions, which mimics the short and strong variability observed in GRBs \cite{1997ApJ...490...92K}. Refs. \cite{Bustamante:2014oka,GAM15a} discussed the effect of the multiple collisions, and demonstrated that different messengers (PeV neutrinos, UHECRs, and gamma-rays) are produced at a different region. Neutrinos are emitted in inner regions with copious target photons, while gamma-rays and UHECRs are produced at outer regions at which they can easily escape without depletion/attenuation. In their calculations, the resulting neutrino fluence is significantly lower than the limits obtained by IceCube. Therefore, with a more realistic situation, the UHECR-GRB paradigm is still viable (see Refs \cite{2017ApJ...837...33B,2020MNRAS.498.5990H} for recent updates). These models can be tested by future neutrino detectors with radio technique \cite{Aartsen:2020fgd}.

\section{Afterglows}\label{sec:afterglow}

\begin{figure}[tb]
 \begin{center}
  \includegraphics[width=\linewidth]{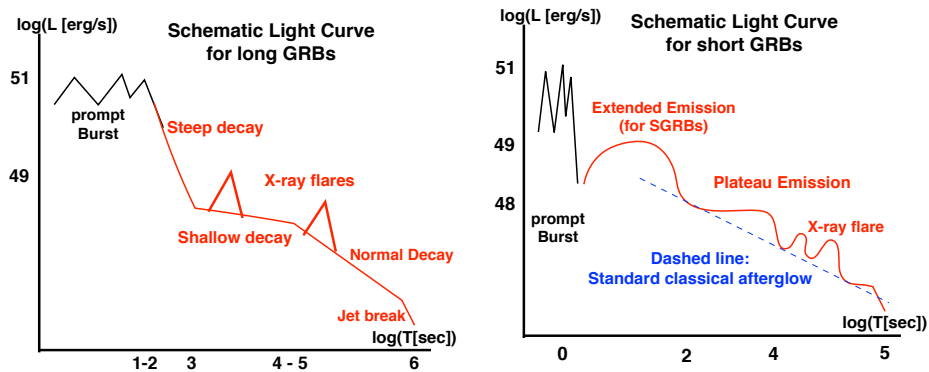}
  \caption{Schematic light curves of prompt and afterglow phases for LGRBs (left) and SGRBs (right). Prompt bursts are followed by the steep decay phase. After that, the shallow decay phase and extended emission phase lasts for $10^4$ and $10^2$ sec for LGRB and SGRBs, respectively. X-ray flares also appear randomly in $10^2 - 10^5$ sec for both LGRBs and SGRBs. For SGRBs, the plateau emission with $T\sim10^4$ sec follows the extended emission. }\label{fig:lightcurves}
 \end{center}
\end{figure}

GRBs are followed by afterglows, broadband transients  with a longer timescale of hours to days. Before the Swift era, the observational data were consistent with the simple forward shock model, in which the afterglow is produced by the forward shock formed by the interaction between the relativistic jets and circum-burst material \cite{MR97a,SPN98a}. The early afterglow observation by Swift revealed that the situation is more complicated than expected. There are many unexpected features in the early afterglow phase (around an hour after the burst), including the steep and shallow decay phases and X-ray flares in LGRBs \cite{2005Sci...309.1833B,GRF09a}, and extended emissions ($\sim10^2$ sec) and plateau emissions ($\sim10^4$ sec) in SGRBs \cite{NB06a,KIS17a} (see Figure \ref{fig:lightcurves}). These features cannot be produced by the standard forward shock emissions \cite{IKZ05a}, and thus, late-time internal dissipation models are proposed in order to explain these features \cite{FW05a,ZFD06a}. This implies that the central engine is active at least for minutes to hours after the burst for SGRBs and LGRBs, respectively. In this section, we briefly discuss neutrino emissions in afterglow phases.

 \begin{figure}[tb]
\begin{center}
  \includegraphics[width=\linewidth]{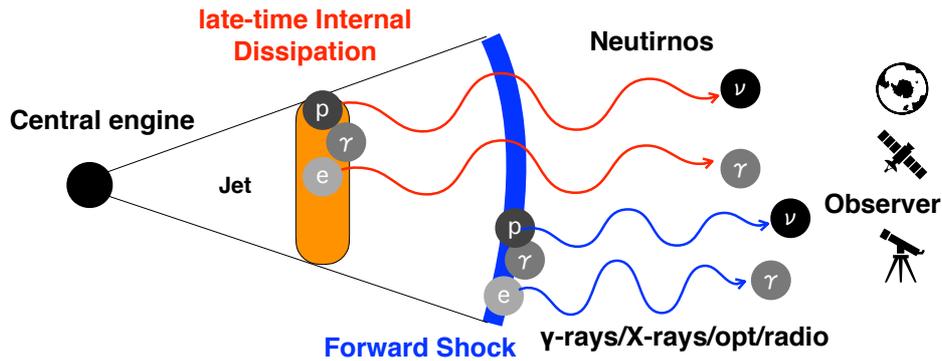}
  \caption{Schematic picture of neutrino emissions from the afterglow phase. Forward shock formed by the interaction between jets and ambient medium accelerate non-thermal electrons that emit broadband photons. Non-thermal protons are also accelerated and produce neutrinos of EeV energies. However, the neutrino production efficiency is so low that it is challenging to detect the neutrinos from the forward shocks. The prolonged central engine activities trigger the internal dissipation in the early afterglow phase, which are potential sites of high-energy neutrino production owing to a small dissipation radius and a low Lorentz factor.}\label{fig:late-time}
 \end{center}
\end{figure}

\subsection{Standard external shock model}
The GRB jets interact with the ambient medium. Since the jet is much faster than the sound speed of the ambient medium, a forward shock propagates in the medium. Non-thermal electrons accelerated in the shock emit broadband photons by the synchrotron process (see Figure \ref{fig:late-time}), which are observed as the afterglow emission \cite{1997Natur.387..783C,1997Natur.386..686V,1997Natur.389..261F}. The forward shock model developed by Refs. \cite{MR97a,SPN98a} predicts a multi-component broken power-law spectrum, and optical and X-ray lightcurves decrease with a power-law manner. This model can reproduce the broadband (radio, optical, X-ray) afterglow observations in a late phase ($T\gtrsim1$ day; see the next subsection for the afterglows in the early phase). High- and very-high-energy gamma-ray emissions via the synchrotron-self Compton process are also predicted \cite{2001ApJ...548..787S}, which is confirmed by the detection of very-high-energy gamma rays from the afterglows of bright GRBs \cite{Acciari:2019dbx,Arakawa:2019cfc}. 

Initially, the jets propagate in the circum-burst medium with a constant velocity, and they are decelerated when the energy of the accumulated material becomes comparable to the jets. The deceleration radius is estimated to be
\begin{equation}
R_{\rm dec}\approx \left(\frac{3E_{K,\rm iso}}{4\pi m_pn_{\rm amb}c^2\Gamma_j^2}\right)^{1/3}\simeq 7.9\times10^{16}E_{K,\rm iso,53.5}^{1/3}\Gamma_{j,2.5}^{-2/3}n_{\rm amb,0}^{-1/3}\rm~cm,
\end{equation}
where $E_{K,\rm iso}$ is the isotropic equivalent kinetic energy of the jets and $n_{\rm amb}$ is the number density of the circum-burst medium.
The lightcurve from the external forward shock peaks at the deceleration time in the observer frame, 
\begin{equation}
 T_{\rm dec} \approx (1+z)\frac{R_{\rm dec}}{\Gamma_j^2 c}\simeq53E_{j,\rm iso,53.5}^{1/3}\Gamma_{j,2.5}^{-8/3}n_{\rm amb,0}^{-1/3}\left(\frac{1+z}{2}\right)\rm~sec 
\end{equation}
Since the emission region of the external forward shock is a few orders of magnitude larger than the internal dissipation region, the neutrino production efficiency is several orders of magnitude lower. The photon spectrum is also softer than the prompt phase, leading to a peak neutrino energy of $E_\nu\sim0.1-10$ EeV \cite{WB00a}. These properties make the neutrino detection more challenging. Indeed, Ref. \cite{2017PhRvD..96j3004T} discussed the detectability of high-energy neutrinos from nearby GRBs. They calculated the neutrino spectra from 23 nearby GRBs using the multi-wavelength data to model the afterglow emission, and found that the neutrino fluence from the nearby GRBs are too low to be detected by a current and near-future detectors, such as IceCube-Gen2 \cite{Aartsen:2019Gen2}.

As the jets interact with the circumburst medium, a reverse shock propagates in the jets. Such a reverse shock can also accelerate particles and emit optical photons, which is observed as optical flashes \cite{1999Natur.398..400A}.
Refs. \cite{WB00a,Mur07a} discussed the neutrino emission from the reverse shocks of the GRB jets. Their predicted neutrino spectra have a peak at the EeV range, and the predicted fluence is lower than the prompt phase or late-time internal dissipation models. Also, optical follow-up observations of GRBs did not detect any signatures from reverse shocks \cite{2007A&A...469L..13M}, which suggest that the particle acceleration at reverse shocks may be inefficient.

\subsection{Late prompt emissions}

XRT onboard Swift discovered mysterious behaviors in X-ray lightcurves of considerable fraction of GRBs. For both LGRBs and SGRBs, the X-ray luminosity suddenly increases in $10^2-10^5$ sec after the prompt bursts, and it sharply drops after its peak luminosity~\cite{2005Sci...309.1833B,2006A&A...454..113C,CMM10a,MCG11a,2016ApJS..224...20Y}. These X-ray flares cannot be explained by the standard forward shock models \cite{IKZ05a}, implying that the central engine is active at such a late phase. In addition, the lightcurves of LGRBs often indicate the plateau emission, or the shallow decay phase, at $T\sim10^2-10^4$ sec \cite{2006ApJ...642..389N}, which can be interpreted as the energy injection by the prolonged central engine activity\cite{ZFD06a}. The lightcurves in SGRBs also have a similar feature, namely extended and plateau emissions~\cite{KIS17a,2019ApJ...877..147K}, which can be attributed to the long-lasting central engine activities for $T\sim10^{2}-10^5$ sec. Since the observed spectrum is softer and the luminosity is lower, the compactness problems for late internal dissipation sites are less serious, and the required values of Lorentz factor for these jets are $\Gamma_j\sim5-50$ \cite{2020MNRAS.493..783M}. In this subsection, we focus on neutrinos from X-ray flares in LGRBs. Neutrinos from prolonged engine activities in SGRBs will be discussed in Section \ref{sec:sgrb}

Origin of the X-ray flares, shallow decay phases, extended emissions, and plateau emissions are not well understood. Since X-ray flares show features similar to the prompt bursts, it is natural to consider internal dissipations as their production mechanism \cite{FW05a,PAZ06a}. Ref. \cite{MN06a} discussed the neutrino emission from the X-ray flares based on the late internal dissipation scenario. They found that the late internal dissipation can produce neutrinos more efficiently owing to their lower Lorentz factors (see e.g., Ref. \cite{2016ApJ...831..111M} for estimate of $\Gamma_j$ for X-ray flares). The X-ray flares have a lower peak photon energy ($E_{\gamma,\rm br}\sim0.1-1$ keV) and a lower Lorentz factor ($\Gamma_j\sim10-100$), although observational support of these values are weak. These values lead to a typical neutrino energy similar to the prompt phase, $E_{\nu,\rm br}\simeq2 \Gamma_{j,1.5}^2(E_{\gamma,\rm br}/1~\rm keV)^{-1} ((1+z)/2)^{-2}$ PeV. 

The predicted diffuse neutrino flux from the X-ray flares can be as high as that from the prompt phase before the IceCube detection  \cite{MN06a}. 
The IceCube GRB analyses mentioned in Section \ref{sec:IceCubeGRB} do not consider contributions from the late internal dissipation, and hence, the neutrino emission from the X-ray flares are currently not constrained. Ref. \cite{2020ApJ...890...83G} analyzed the publicly available data of IceCube experiment taking into account the delayed internal dissipation, and found a hint of neutrino signal about a day after the burst. Although the statistical significance is not high, future dedicated searches will shed light on the prolonged engine activity and constrain the parameters of the mysterious phenomena.

\section{LLGRBs and choked jets} \label{sec:llgrb}

The luminosity of GRB 980425, the first LLGRB, is $5\times10^{46}\rm~erg~s^{-1}$, which is $10^3-10^5$ times lower than that for a typical GRB. LLGRBs have observational features different from typical luminous LGRBs, in addition to the low luminosity. LLGRBs tend to have smooth lightcurves, compared to those for luminous GRBs. The typical duration of LLGRBs is $10^2-10^3$ sec, which is slightly longer than that for a typical LGRB. LLGRBs do not follow the $L_{\gamma,\rm iso}-E_{\gamma,\rm br}$ (Yonetoku) relation \cite{2004ApJ...609..935Y,2010PASJ...62.1495Y}. Also, the luminosity function of the observed LLGRBs is higher than the extrapolation of the luminous GRBs \cite{2007ApJ...662.1111L}. Ultra-long GRBs are also another sub-class of low-power GRBs. Their luminosity is comparable to the LLGRBs, but the duration of ultra-long GRBs are much longer, typically $\sim10^4$ sec\cite{2014ApJ...781...13L}.

These distinctive features suggest a different emission mechanism for LLGRBs. Two popular scenarios are actively discussed. One is the emission from low-power and slower ($\Gamma_j\lesssim30$) jets \cite{1999ApJ...516..788W}, also known as baryon-rich jets. Owing to their lower power and lower $E_{\gamma,\rm br}$, LLGRBs are allowed to have lower Lorentz factor jets.
The cosmic-ray and neutrino emissions in the slower jet scenario is discussed in Section \ref{sec:slowjets}. 
%Current LLGRB data is also consistent with the extrapolation of the faint end of the luminous GRB luminosity function \cite{2015ApJ...812...33S}. 

The other is the shock breakout scenario, in which jets can be choked inside the progenitor star or the dense stellar wind. The shock formed by the jet-progenitor star interaction keeps propagating the surrounding material, and eventually breaks out from the surface, which is observed as LLGRBs. If neither the jets nor the shocks penetrate the progenitor star, the jet kinetic energy is stored in the stellar material, which results in a SN with a higher kinetic energy~\cite{1998Natur.395..663K}. This can be observed as a trans-relativistic SN, which may be regarded as a choked-jet powered SN.  We will discuss this type in Section \ref{sec:choked}.

\subsection{Baryon-rich jet scenario}\label{sec:slowjets}

LLGRBs can be a sub-class of GRBs with low Lorentz factor and low power jets \cite{1999ApJ...516..788W}. In this context, the emission mechanisms of LLGRBs can be internal dissipation, and the formalism discussed in Section \ref{sec:prompt} is applicable with an appropriate parameter choice. The local event rate and characteristic luminosity of LLGRBs is estimated to be $\rho_{\rm LLGRB}\sim100-400\rm~Gpc^{-3}~yr^{-1}$, and the characteristic luminosity is $L_{\gamma,\rm iso}\sim10^{47}\rm~erg~s^{-1}$ (corresponding to $\mathcal{E}_{\gamma,\rm iso}\sim10^{50}$ erg with a duration of $\sim10^3$ sec) \cite{2007ApJ...662.1111L,2015ApJ...812...33S}. Since the energy budget by LLGRBs is about an order of magnitude lower than that by LGRBs, the baryon loading factor should be $\xi_p\sim100 f_{\rm bol,-1}$ in order to explain the UHECR data \cite{mur+06,ZMK18a}. This value is still acceptable. $f_{\rm bol}\sim1-2$ can be achieved with a hard proton spectral index of $s\lesssim1.5$, which makes $\xi_p\sim10-20$. Refs. \cite{mur+06,gz07} discussed the neutrino emissions from LLGRBs. They found that the contribution to the diffuse neutrino flux by LLGRBs can dominate over that by luminous GRBs with an optimistic parameters. Although a single LLGRB cannot be detected, the higher event rate enables LLGRBs to be an important source of the diffuse neutrino flux.

Pierre Auger Observatory reported that the composition of UHECRs is heavier for a higher energy~\cite{Aab:2014kda}, and the combined fit of the airshower maximum depth and the UHECR spectrum suggests that the UHECR sources need to have a chemical composition much heavier than the solar metalicity. The chemical composition of the progenitor star of GRBs are known to have a heavier composition. Ref. \cite{ZMK18a} estimated the chemical composition of LLGRB jets using the prgenitor star model of GRBs \cite{WH12a}, and demonstrated that LLGRBs can naturally account for the observed UHECR composition and spectrum simultaneously. Since heavy nuclei are likely destroyed via the photo-disintegration in luminous GRBs because of their dense radiation fields, LLGRBs are a better candidate of the source of UHECR nuclei \cite{HMI12a,ZMK18a}.

In the dissipation region of LLGRBs, neutrinos are also produced because some fraction of UHECRs interact with the LLGRB photons. The typical energy of the neutrinos is estimated to be $E_{\nu,\rm br}\simeq 0.70\Gamma_{j,1}^2(E_{\gamma,\rm br}/1{\rm~keV})^{-1}(1+z)^{-2}$ PeV. The estimated contribution by LLGRBs to cosmic high-energy neutrino background is comparable to the IceCube diffuse neutrino flux~\cite{ZMK18a}. Ref. \cite{2019ApJ...872..110B} demonstrated that LLGRBs can account for both Auger and IceCube data by searching a wide parameter range of dissipation radius and luminosity.

\subsection{Choked jet scenario}\label{sec:choked}

 \begin{figure}[tb]
\begin{center}
  \includegraphics[width=\linewidth]{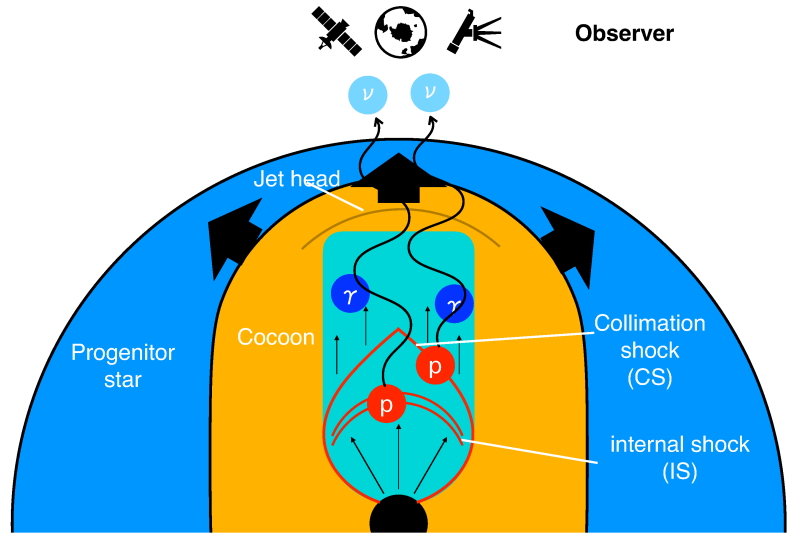}
  \caption{Schematic picture of choked jet systems. Jets interact with the progenitor star, which forms a jet head structure consisting of forward and reverse shocks. The shocked material expands sideways, and create a cocoon surrounding the jets. The cocoon pushes the jets inward and collimate them, causing the formation of the collimation shock. Fluctuations of the jet velocity can result in internal shocks in un-collimated jets..}\label{fig:choked}
 \end{center}
\end{figure}

When the progenitor star of the LLGRB collapses, relativistic jets are launched from the central engine as is the case for luminous GRBs. The jets interact with the progenitor star, and propagates in the surrounding medium. They may fail to penetrate the progenitor star, i.e., jets are choked, if the duration of the jet production is short or if the progenitor star has an extended envelope. When the jet quenching is close to the photosphere of the progenitor, shocks can break out from the surface, which produces a burst emission in X-ray to MeV gamma-ray bands. This mechanism can naturally explains the duration, peak energy, and luminosity of the observed LLGRBs  \citep{WMC07a,NS12a}. The schematic picture of the choked jet system is illustrated in Figure \ref{fig:choked}. In this section, we focus on the choked jet system in the collapsar scenario. It is also possible that BNS mergers lead to choked jet systems, which is discussed in Section \ref{sec:bns-choked}

\subsubsection{Successful jets and choked jets}

Since the jet propagation velocity is much higher than the sound velocity in both the jets and the stellar envelope, the interaction between the jets and the envelope forms the forward and reverse shocks propagating in the progenitor star and in the jets, respectively. The interaction region is called the jet head, whose propagation velocity is determined by balancing the ram pressures in the jet and stellar material at the jet-head rest frame \cite{2003MNRAS.345..575M,BNP11a}:
\begin{equation}
 \rho_j h_j \Gamma_j^2\Gamma_h^2(\beta_j - \beta_h)^2c^2 + P_j = \rho_a h_a \Gamma_h^2\beta_h^2c^2 + P_a,
\end{equation}
where  $\rho_i$,  $P_i$, $h_i=1+4P_i/(\rho_i c^2)$ are the density, pressure, enthalpy measured in the fluid rest frame, respectively. $\Gamma_i$ and $\beta_i$ are the Lorentz factor and the velocity measured in the rest frame of the ambient medium, which is the same as the engine frame. The subscripts $j$, $a$, and $h$ indicates the quantities in the jet, ambient medium, and jet head, respectively. For the case with $P_j\ll \rho_j c^2$ and $P_a\ll \rho_a c^2$, we can write 
\begin{equation}
 \beta_h\approx \frac{\beta_j}{1+\tilde{L}^{-1/2}},
\end{equation}
\begin{equation}
 \tilde{L}=\frac{\Gamma_j^2\rho_jh_j}{\rho_a}\approx \frac{L_j}{A_j\rho_ac^3},
\end{equation}
where $L_j =\theta_j^2 L_{j,\rm iso}/4$ is the intrinsic jet kinetic luminosity, $\theta_j$ is the jet opening angle, and $A_j$ is the cross section of the jet and ambient medium. They are related to the jet quantities by $L_j\approx \Gamma_j^2 \rho_jh_j A_j c^3\beta_j$, and we use $\beta_j\sim1$. From these equations, we see $\beta_h\sim\beta_j\sim1$ for $\tilde{L}\gg1$ and $\beta_h\sim\tilde{L}^{1/2}$ for $\tilde{L}\ll1$. The density for a Wolf-Rayet star of mass $M_*$ and radius $R_*$ is $\rho_a \approx 3M_*/(4\pi R_*^3)\sim4.8M_{*,1}R_{*,11}^{-3}\rm~g~cm^{-3}$, where  $M_{*,1}=M_*/(10~M_\odot)$. Then, $\tilde{L}\sim2\times10^{-3}L_{j,50}\theta_{j,-1}^{-2}M_{*,1}^{-1}R_{*,11}\ll1$, where we use $A_j=\pi \theta_j^2R_*^2$. Thus, the jet head propagates in the stellar material with a non-relativistic speed.

The shocked matters for both stellar and jet materials expand sideways, forming a cocoon surrounding the jets. If $\tilde{L}\gg 1$, i.e., $\beta_h\sim\beta_j$, the jet energy is spent to the expansion motion. The cocoon pressure is so weak that it cannot collimate the jets. On the contrary, if $\tilde{L}\ll1$, namely $\beta_h < 1$, most of the jet power is stored in the cocoon. The cocoon pushes the jet inward, and jets are easily collimated. See Ref. \cite{BNP11a} for more accurate conditions for the jet collimation. Collimation changes the jet cross section, $A_j$, and then, we need to formulate the jet propagation consistently including the collimation effect in order to estimate the jet breakout time. 
Appropriately taking the collimation effects into account, the typical timescale for jets to penetrate a Wolf-Rayet star is estimated to be $t_{\rm brk}\sim15$ sec \cite{BNP11a,mi13}.

This breakout time is supported by the observed duration distribution of the GRBs \cite{2012ApJ...749..110B}. The observed GRB duration, $t_\gamma$, can be represented by $t_\gamma= t_{\rm eng} - t_{\rm brk}$, where $t_{\rm eng}$ is the duration of the prompt engine activity. Suppose that the probability distribution of $t_{\rm eng}$, $p_e(t_{\rm eng})$, is a smooth function of time. Then, the observed duration distribution, $p_\gamma(t_\gamma)$, is given by $p_\gamma(t_\gamma)\approx p_e(t_{\rm eng}=t_\gamma)$ for $t_\gamma\gg t_{\rm brk}$ and $p_\gamma(t_\gamma)\approx p_e(t_{\rm eng}=t_{\rm brk})=\rm const$ for $t_\gamma < t_{\rm brk}$. Indeed, the observed duration distributions by BATSE, Swift-BAT, and Fermi-GBM have a plateau feature in $T\lesssim20-30$ sec, which supports the collapsar scenario with $t_{\rm brk}\sim20-30$ sec. 
The GRB duration distributions for $t_\gamma> t_{\rm brk}$ can be fitted by a power-law function: $p_\gamma(t_\gamma \gg t_{\rm brk})\propto t_\gamma^{-\alpha}$ with $\alpha\approx3-4$. This implies that the choked-jet population, i.e., collapsars with $t_{\rm eng}<t_{\rm brk}$, dominates over the successful GRBs in terms of the event rate. Indeed, the observed LLGRB rate can be much higher than the LGRB rate\cite{2007ApJ...662.1111L}, which supports the shock breakout emission from choked jets as the origin of LLGRBs.

\subsubsection{Shocks in choked jets and radiation mediated constraints}\label{sec:radiation-mediated}

In the choked jet scenario, photons emitted in the jets are completely absorbed by the stellar material, while the neutrinos can arrive at the Earth because the stellar material is transparent to them. Therefore, the neutrinos are a unique messenger to probe the engine activities and jets in the choked jet systems.
There are four shocks in the choked jet system. Two are the forward and reverse shocks in the jet head. Another is the collimation shock in the collimated jet. The other is the internal shock in the uncollimated jet (see Figure \ref{fig:choked}). 

The choked jet system is very compact and dense environment, where the radiation pressure dominates over the thermal pressure at the shock downstream. If the shock upstream is optically thick to Thomson scattering, the photons leaked from the downstream decelerates the upstream material. This results in radiation mediated shocks, rather than the collisionless shocks mediated by plasma-instabilities \cite{2020PhR...866....1L}.  The size of the radiation-mediated shock is as large as the Thomson mean free path, $l_T\sim 1/(n_e\sigma_T)\sim 2\times10^4 (n_e/10^{20}\rm~cm^{-3})^{-1}$ cm. This scale is much longer than the plasma skin depth, $\lambda_{p,\rm pl}=c/\omega_{p,\rm pl}\approx 2\times10^{-3} (n_i/10^{20}\rm~cm^{-3})^{-1/2}$, where $\omega_{p,\rm pl}=\sqrt{4\pi n_i e^2/m_i}$ is the proton plasma frequency. The mean-free path of the plasma particles are likely comparable to the plasma skin depth, and thus, the plasma particles cannot travel between the shock upstream and down stream if the shock is mediated by radiation. Therefore, radiation-mediated shocks cannot accelerate particles by the diffusive shock acceleration mechanisms. Other acceleration mechanisms are also unlikely because strong radiation fields are expected to suppress the turbulence and magnetic reconnections.

The condition for radiation mediated shocks is given by $\tau_u=n_u\sigma_Tl_u \lesssim 1$ for non-relativistic shocks and $\tau_u\lesssim 0.1\Gamma_{\rm sh}/(1+2\ln\Gamma_{\rm sh}^2)$ for ultra-relativistic shocks \cite{NS12a}. The modification in ultra-relativistic regime is due to the efficient pair production in the relativistic shock.  For the case with the choked jet systems, the shock velocity can be mildly relativistic, and then, the exact condition is in between the two regime. Ref. \cite{mi13} discussed the radiation mediated conditions for the choked jet systems, and concluded that radiation mediated conditions cannot be avoided inside the Wolf-Rayet stars unless we consider extremely efficient jet acceleration up to $\Gamma_j>10^3$ for a typical luminous GRBs. On the other hand, if we consider a low-power GRB jets, such as ultra-long GRBs or LLGRBs, radiation mediated conditions can be avoided for collimation shocks and internal shocks. Ref. \cite{SMM16a} extended their analysis for dense stellar winds, and showed that the radiation mediated conditions can be avoided if the jets dissipates their energy in the extended stellar winds. On the other hand, shocks at the jet head is so dense that they are unlikely to accelerate cosmic-ray particles. The radiation-mediated conditions were often overlooked, and several papers discussed the neutrino emission from the choked jets without evaluating the radiation-mediated condition \cite{MW01a,2003PhRvD..68h3001R,2005PhRvL..95f1103A,HA08a}.

\subsubsection{Neutrino emission from choked jets}

\begin{figure}[tb]
\begin{center}
  \includegraphics[width=\linewidth]{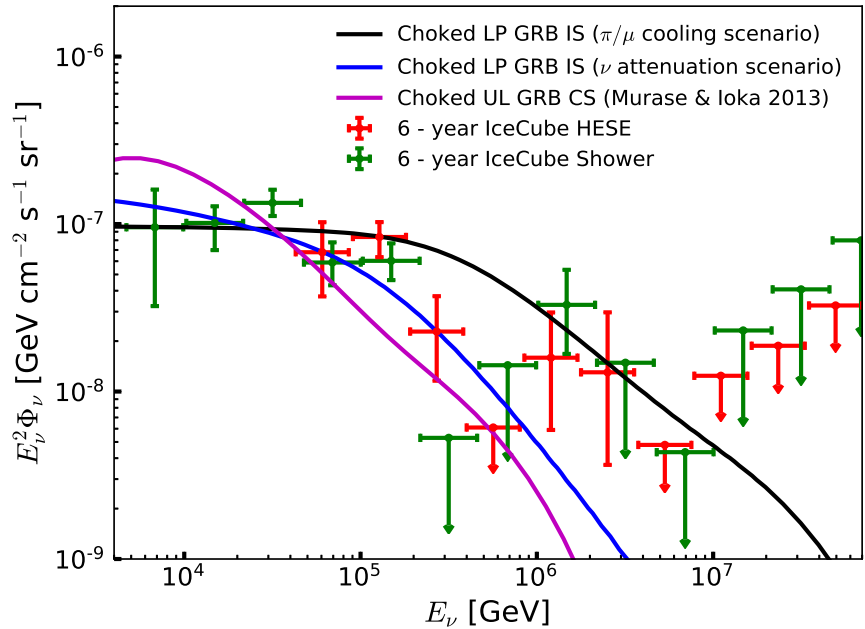}
  \caption{Predicted diffuse neutrino intensities from low-power GRBs with choked jets. Solid lines are theoretical predictions of diffuse neutrino intensities from low-power GRBs (both internal shocks and collimation shocks). The points with error bars represent the IceCube data by shower and High-energy starting event analyses.  Reproduced from Ref. \cite{2020PhRvD.101l3002C} with permission by APS.  }\label{fig:diffuse_choked}
 \end{center}
\end{figure}

Since the choked jets are compact systems with high-power jets compared to other astrophysical environments, the neutrino production there has been intensively discussed \cite{MW01a,2003PhRvD..68h3001R,2005PhRvL..95f1103A,HA08a,mi13,SMM16a,HKN18a}. If we consider the energy dissipation in the stellar envelope of $R\sim10^{11}-10^{12}$ cm, the dissipation radius is much smaller than the canonical GRB internal shocks. Also, there are copious target photons provided from the stellar envelope or stalled jets in which photons are thermalized. Hence, the neutrino production efficiency is higher than the canonical GRB internal dissipations, and usually the system is calorimetric, i.e., all the accelerated protons are depleted by the photomeson production. As suggested by the duration distribution, the event rate of the choked jets are likely more abundant than LLGRBs, and the local rate of choked jet events can be written as $\rho_{\rm cj} = f_{\rm cj}\rho_{\rm HN}$, where $f_{\rm cj}$ is a parameter and $\rho_{\rm HN}\sim5\times10^3\rm~Gpc^{-3}~yr^{-1}$ is the event rate of hypernovae \cite{2011MNRAS.412.1522S,2017PASP..129e4201S}. Then, the CR energy budget of the choked jet system is estimated to be $\mathcal{E}_p\rho_{\rm cj}\simeq 1.6\times10^{45}\mathcal{E}_{p,51}f_{\rm cj,-0.5} \rm~erg~yr^{-1}~Mpc^{-3}$. The all-flavor diffuse neutrino intensity from the choked jet systems is estimated to be 
\begin{eqnarray}
& &E_\nu^2\Phi_\nu \approx \frac{c}{4\pi H_0}\frac38f_zf_{\rm bol}\mathcal{E}_p\rho_{\rm cj}\\
& &\sim1\times10^{-7}{\rm~GeV~s^{-1}~cm^{-2}~sr^{-1}}f_{\rm bol,-1}\left(\frac{f_z}{3}\right)\left(\frac{\rho_{\rm cj}\mathcal{E}_p}{1.6\times10^{45}\rm~erg~Mpc^{-3}~yr^{-1}}\right).\nonumber
\end{eqnarray}
This value is about an order of magnitude higher than that by the classical GRB model given in Section \ref{sec:prompt}. 
Although uncertainties in the choked jet system is large because of the lack of the observational data, the choked jet scenario can explain the cosmic high-energy neutrino background detected by IceCube (see Figure \ref{fig:diffuse_choked}).

Interestingly, the choked jet system can be the source of the mysterious ``medium energy excess'' in the cosmic high-energy neutrino background at the 10-100 TeV range \cite{Aartsen:2020aqd}. The neutrino intensity of the medium energy excess, $E_\nu^2\Phi_\nu\sim1\times10^{-7}\rm~GeV~cm^{-2}~s^{-1}~sr^{-1}$, is higher than the isotropic cosmic gamma-ray background at the sub-TeV range, $E_\gamma^2\Phi_\gamma\sim2\times10^{-8}\rm~GeV~cm^{-2}~s^{-1}~sr^{-1}$ \cite{Ackermann:2014usa}. Therefore, the neutrino source of the medium energy excess should be hidden in gamma-rays, otherwise the gamma-rays accompanied by the neutrinos overshoot the Fermi data \cite{Murase:2015xka}. In the choked jet systems, all the gamma-rays emitted from jets are absorbed by the progenitor star or the dense stellar winds, and thus, they are regarded as hidden neutrino sources.  Only a handful sources can explain the medium-energy excess, making the LLGRB and choked jets an attractive neutrino source. Another candidate for the origin of the medium-energy excess is accretion flows in active galactic nuclei \citep{kmt15,Murase:2019vdl,2019ApJ...880...40I}.

\subsubsection{Observational constraints on choked jets}

The choked jet scenarios can be explored and constrained by searches of neutrinos associated with a sub-class of SNe. Since photons from the choked jets are completely absorbed by the stellar and wind material, neutrinos are the unique signal to probe the high-energy phenomena in the choked jet systems. Ref. \cite{DT18a} constructed a model that describes both successful-GRB and choked-jet populations, and calculated the neutrino flux from them. They found that the diffuse neutrino flux from choked-jet component overshoots the observed IceCube data if the event rate of the choked-jet SNe is higher than 1\% of the normal core-collapse supernoevae. Ref. \cite{2018JCAP...01..025S} and Ref. \cite{2018JCAP...12..008E} searched for neutrino events associated with the envelope-stripped SNe using the 1-year track events  and 6-year high-energy starting events of IceCube, respectively. Both analyses found no associated neutrino, and put constraints on the choked-jet fraction to all the stripped-envelope SNe. Although their constraints are not significant with their data sets, they demonstrated that the methods would put a meaningful constraint on the choked-jet SN fraction near future with better facilities of optical transient surveys (e.g., Vela Rubin Observatory \cite{LSST09a}) and neutrino telescopes (e.g., IceCube-Gen2 \cite{Aartsen:2020fgd} and KM3NeT \cite{KM3NeT16a}), as discussed in Ref. \cite{2020MNRAS.492..843G}.

\section{Neutrinos from BNS mergers}\label{sec:bns}

SGRBs are a sub-class of GRBs, whose duration is less than 2 seconds. The event rate of SGRBs is estimated to be $\rho_{\rm SGRB}\sim4-10\rm~Gpc^{-3}~yr^{-1}$\cite{NGF06a,WP15a}, which is higher than that of LGRBs. However, SGRBs are less energetic, $\mathcal{E}_{\gamma,\rm iso}\sim10^{51}$ erg\cite{Nak07a}. The energy budget of the SGRBs is at most $\rho_{\rm SGRB}\mathcal{E}_{\gamma,\rm iso}\sim1\times10^{43}\rm~erg~Mpc^{-3}~yr^{-1}$, which is an order of magnitude lower than that of LGRBs. Thus, the SGRBs contribution to the cosmic high-energy neutrino background is not important, compared to LGRBs.

Nevertheless, SGRBs are interesting targets of multi-messenger astrophysics. They are believed to occur when a BNS merges \cite{1986ApJ...308L..43P,NPP92a}. This scenario is strongly supported by the multi-messenger event, GW 170817 \cite{LIGO17c,LIGO17d,LIGO17e}. In 2017, the LIGO/VIRGO Collaboration detected the gravitational waves from a BNS merger. About 2 second later, Fermi and INTEGRAL satellites detected a short gamma-ray burst lasting about 2 seconds. Roughly half a day later, optical/ultraviolet/infrared telescopes reported a discovery of a transient object in a galaxy located at 40 Mpc away from the Earth. A gradually brightening object was found in  X-ray and radio bands 9 and 11 days after the merger event, respectively. VLBI observations and afterglow lightcurve modelings evidenced that the observed X-ray and radio signals originated in a collimated jet seen from an off-axis observer \cite{MDG18a,2018MNRAS.478L..18T}. High-energy gamma-rays and high-energy neutrinos were not detected \cite{IceCube17c,HESS17a,Fermi18a}, although they are expected to be produced in SGRBs \cite{2001ApJ...548..787S,2009ApJ...707.1404T,MTF18a,2019ApJ...887L..16K}. 

The localization of the gravitational wave events are typically $\sim10-100\rm~deg^2$, which is much larger than the typical field of view of the follow-up facilities in radio, optical, and X-ray bands. High-energy neutrino detectors are able to monitor all the sky all the time, and thus, the high-energy neutrino signals are suitable for multi-messenger connection with gravitational wave transients. In this section, we discuss the prospects of high-energy neutrino detection coincident with gravitational waves.

\subsection{Late-time internal dissipation scenario}\label{sec:sgrb}

 \begin{figure}[tb]
\begin{center}
  \includegraphics[width=\linewidth]{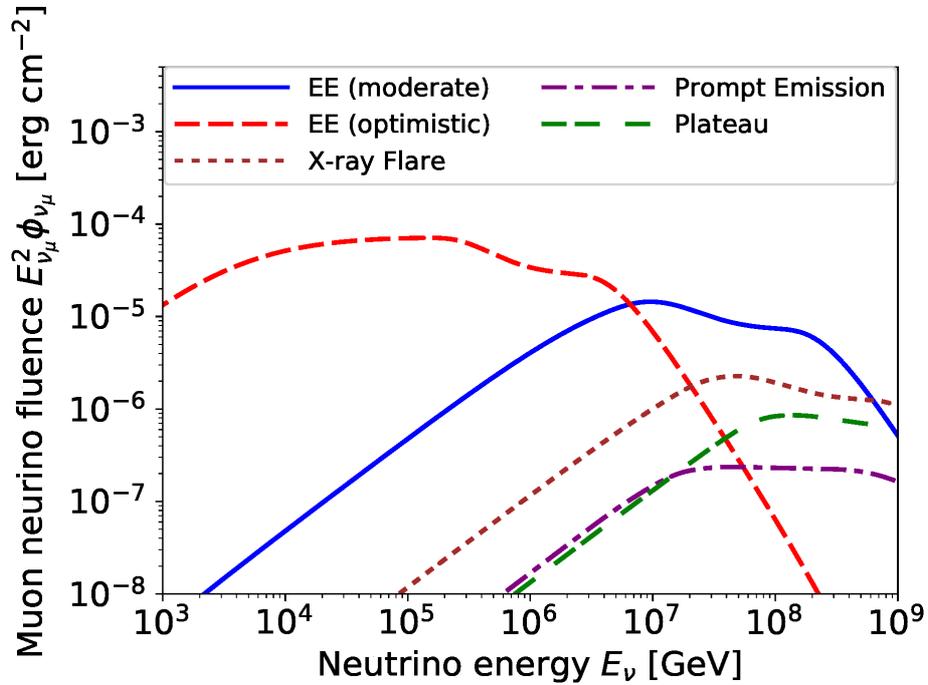}
  \caption{Differential neutrino fluences from various components of internal dissipation regions in SGRBs. We see that the extended emission component is the most efficient neutrino emitter. Reproduced from Ref. \cite{KMM17b} with permission by AAS.}\label{fig:sgrb}
 \end{center}
\end{figure}

A significant fraction of SGRBs have extended emission and plateau emission components in their X-ray lightcurves \cite{KIS17a,2019ApJ...877..147K}, which are likely produced at late-time internal dissipation regions by the prolonged central engine activities (see Figure \ref{fig:late-time}). Characteristics of the late-time internal dissipation regions are not well constrained, and they can be an efficient neutrino emitter. We expect gravitational wave signals from nearby BNS merger events, and coincident searches for gravitational waves and neutrinos are powerful tool to probe the prolonged central engine activity.
%The contribution by SGRBs to the cosmic neutrino background are sub-dominant because LGRBs totally dominates in the view point of energetics. Nevertheless, it is possible to detect transient high-energy neutrino signals from late-time internal dissipation in SGRBs, if a nearby SGRB occurs. 

Ref. \cite{KMM17b} discussed prospects for coincident detection of neutrinos and gravitational waves. They calculated neutrino spectra from various components, namely prompt bursts, extended emissions, plateau emissions, and X-ray flares. They found that the extended emission components can emit high-energy neutrinos most efficiently (see Figure \ref{fig:sgrb}). Based on the calculated spectra and approximately taking into account the distribution of Lorentz factor, prospects of high-energy neutrino detection coincident with gravitational waves are discussed. For an optimistic scenario, the coincident detection with IceCube is probable, and IceCube-Gen2 will likely detect high-energy neutrinos even with a moderate parameter set. Thus, neutrino observations will be able to put a meaningful constraint on the parameter space of prolonged central engine activities with a current and planned facilities.

Interestingly, the Lorentz factors used for the extended emission components, $\Gamma_j=10$ and 30 for optimistic and moderate cases, respectively, are consistent with the estimates by a later study \cite{2020MNRAS.493..783M}. $\Gamma_j\sim10$ is supported by the compactness arguments against the extended emission component. $\Gamma_j\sim30$ is plausible based on a scenario where the plateau emission component is powered by the forward shock energized by the jets of the extended emission component.

\subsection{Choked jet scenario}\label{sec:bns-choked}

In the BNS merger paradigm, SGRBs were first considered to occur in a ``clean'' environment in the sense that the jet does not interact with the ambient medium. However, numerical simulations revealed that the BNS mergers eject a copious amount of neutron rich material of $M_{\rm ej}\sim10^{-3}-10^{-2}$ \cite{HKK13a,BGJ13a}. The rapid neutron capture nucleosynthesis process (r-process)  creates neutron-rich nuclei inside the ejecta, which results in an optical/infrared transient powered by the nuclear decay of neutron-rich nuclei \cite{LP98a,MMD10a,BK13a,TH13a}. Indeed, a kilonova was observed as an optical/infrared counterpart of GW170817 \cite{CFK17a,MASTER17a,DES17a,DLT4017a,JGEM17a}. In addition, a few kilonova candidates are reported in the afterglows of nearby SGRBs \cite{2013Natur.500..547T,Troja+18-sgrb150101Bkn,2019ApJ...883...48L}.

Regarding the GW-GRB connection of the GW 170817 event, the delay time of the gamma-ray signal relative to the gravitational-wave signal is about 2 second. A part of the delay can be attributed to the time lag between the merger event and jet launch~\cite{GNP18a,2019ApJ...877L..40G,2020MNRAS.491.3192H}. The time lag of 1 second results in choked jets in the context of SGRBs (see Figure \ref{fig:choked}), as demonstrated in Ref. \cite{MK18a} by semi-analytic calculations and in Ref.\cite{GNP18a} by numerical simulations. In addition, the duration distribution of SGRBs supports existence of choked-jet systems in SGRBs \cite{2017MNRAS.472L..55M}.

Ref. \cite{KMB18a} discussed the neutrino production from the choked jet systems in the context of BNS mergers and SGRBs. From this system, gamma rays from the choked jets cannot escape from the system because of the absorption by the kilonova ejecta, and thus, neutrinos are a unique high-energy counterpart to gravitational waves. The radiation mediated conditions (see Section \ref{sec:radiation-mediated}) can be avoided even for the choked jets in BNS mergers with $\Gamma_j\gtrsim200$ for internal shocks and $\Gamma_j\gtrsim500$ for collimation shocks. Pion coolings are very effective in the collimation shocks, which completely suppress the neutrinos of $\gtrsim1$ TeV, making the neutrino detection very challenging. On the other hand, the internal shocks can produce neutrinos of $E_\nu\sim1-100$ TeV, enabling the neutrino detection coincident with gravitational waves by IceCube in a few years of operation with an optimistic parameters and the design sensitivity of the second generation gravitational wave detectors. Future multi-messenger observations of neutrinos and gravitational waves will shed light on the central engine activities in SGRBs and BNS mergers.

\subsection{Remnant compact object scenario}

BNS mergers leave a rapidly rotating NS or BH as a merger remnant.  Both of them can produce powerful outflows that power electromagnetic and neutrino transients. These scenarios are particularly important for gravitational-wave counterparts because the neutrinos are emitted almost isotropically. The neutrinos from the jets discussed in the previous sections are beamed, and thus, the apparent event rate is lower by a factor of $\theta_j^2$ than isotropic emission scenarios.

A rapidly rotating NS produces a powerful relativistic wind using its spindown power, as in pulsars. The wind collides with the ejecta of the kilonova, and non-thermal electrons are accelerated at the reverse shock in the wind. They emit broadband photons through synchrotron and inverse Compton processes, shining as a merger-remnant wind nebula. These photons are detectable after the kilonova ejecta becomes transparent to the relevant wavelength, typically a week for hard X-rays/GeV gamma-rays and  a month for soft X-rays \cite{MTF18a}. Current facilities can detect these signals, otherwise we can put meaningful constraints on the newborn NS parameters. If non-thermal protons are also accelerated in the nebula, the protons interact with the nebula photons, leading to an efficient high-energy neutrino production \cite{2017ApJ...849..153F}. In this scenario, pions are efficiently cooled at the initial stage of $t<10^4$ sec, while the pion production becomes less efficient for $t>10^6$ sec. Thus, high-energy neutrinos are most efficiently produced around a day after the merger. The neutrino spectrum has a peak at 0.1--1 EeV if we assume that the target photons are dominated by emission from the thermal electrons, which will be testable by future experiments, such as IceCube-Gen2 \cite{Aartsen:2020fgd} and GRAND \cite{2020SCPMA..6319501A}. 

A remnant BH can also produce outflows through the fallback accretion, which may form the fallback-driven outflow nebula. Fallback accretion luminosity decreases faster than the neutron-star spindown luminosity, and thus, it is challenging to detect the high-energy emissions from the fallback driven outflow nebula. Radio signals may be detectable with an optimistic situation \cite{MTF18a}. The fallback-driven outflows contain baryons, and CR protons are likely accelerated there. Ref. \cite{2020JCAP...04..045D} discussed the neutrino emissions from the fallback-driven outflows, and found that neutrino detection from the fallback-driven outflows are challenging even with the planned detectors.

\subsection{Current neutrino observations and constraints}

IceCube, ANTARES, and Pierre Auger Observatory searched for high-energy neutrino signals associated with gravitational waves. Since the binary BHs are unlikely to produce GRBs, we focus on the BNS merger case here. Based on the first gravitational wave transient catalog, we have 9 binary BH mergers and a BNS merger, GW170817\cite{LIGOScientific:2018mvr}. No high-energy neutrino is detected from the multi-messenger event, GW170817, despite that the event is exceptionally nearby (40 Mpc; typical SGRBs occur at $\sim1$ Gpc). The upper limit obtained by Ref. \cite{IceCube17c} is below the prediction for the optimistic extended emission component given by Ref. \cite{KMM17b}. Nevertheless, we cannot put meaningful constraint on the jet parameters by this event. According to the afterglow lightcurve modelings, the jet axis of GW170817 is misaligned to the Earth. This makes the neutrino fluence much lower than that for an on-axis observer, resulting in the theoretical prediction much below the upper limit. Ref. \cite{BHW18a} estimated the neutrino fluence in the prompt phase of GW 170817 using the gamma-ray data, and found that non-detection is consistent with the prediction by the internal dissipation scenario. Future analyses with a larger sample size or an on-axis event will put meaningful constraints on the physical conditions of the jets and the remnant objects.

\section{Summary \& Outlook}

GRBs have been actively discussed as a prime candidate of UHECR and astrophysical high-energy neutrino sources for a long time, especially since they turned out to be explosions at a cosmological distance. GRBs are energetically reasonable as a source of UHECRs, because the required UHECR luminosity is comparable to the gamma-ray luminosity. Based on the standard internal shock scenario, electrons accelerated at the shock emit the observed gamma-rays via synchrotron radiation. If protons are accelerated simultaneously, they should produce copious neutrinos of PeV energies by interactions with the gamma rays.
Despite the high expectation of neutrino detection, IceCube found no neutrino associated with electromagnetically detected GRBs so far, which ruled out the UHECR-GRB connection by the internal shock scenario with the simple one-zone approximation. The multiple shell collision scenario is proposed as a more realistic alternative. In this scenario neutrinos are efficiently produced in the inner part of dissipation regions, while UHECRs and gamma-rays are produced at the outer part. This scenario can explain current UHECR data without violating the neutrino data, and will be tested by future experiments with radio techniques.

GRBs are followed by afterglows, a broadband emissions of a longer timescale. In the standard scenario, afterglows are emitted by the electrons accelerated at the external forward shock. Protons may be accelerated simultaneously, but the emission region is so large that we cannot expect the efficient neutrino production. It is challenging to detect neutrinos from the external forward shocks near future.
Swift-XRT revealed that the X-ray flares and shallow-decay phases are common in the early afterglow phase for LGRBs ($T\lesssim10^4$ sec). These components cannot be attributed to the standard forward shock scenario, implying that the central engine should be active for a longer time than expected. If the late-time internal dissipation powers X-ray flares, the neutrinos should be produced with the same mechanism as in the prompt phase. Owing to their lower Lorentz factor, X-ray flares may be an efficient neutrino emitter. Current IceCube analyses do not constrain signals from the late-internal dissipation scenario. A dedicated analysis is necessary to obtain a tight constraint on the neutrino emission from the X-ray flares.
%However, the energy fluence of X-ray flares are at most comparable to the prompt radiation of the GRBs, and its softer spectrum makes the typical neutrino energy higher. Thus, the X-ray flares are unlikely to be the source of the cosmic high-energy neutrino background, 

LLGRBs and choked jet systems can be distinct populations from canonical GRBs. Although the parameters of the system are only weakly constrained from the observations because of their low luminoisties, LLGRBs and choked jet systems are possible to explain the bulk of the diffuse high-energy neutrino flux observed by IceCube including the medium energy excess. The choked jet system should accompany a powerful SN. Using the optical transient data, we can explore the parameter space of the choked jet system. The constraints with the current data is not significant, but we will be able to put a meaningful constraint near future owing to the better optical transient search programs and the better neutrino detectors in both hemispheres.

Gravitational wave detection by LIGO/VIRGO Collaborations provided us a new window to probe the high-energy phenomena in the Universe. From the multi-messenger event, GW 170817, BNS mergers are supported to be the progenitor of SGRBs. High-energy neutrinos are efficiently produced in the late-time internal dissipation regions,  choked jets, and remnant-powered wind nebulae. The coincident detection of high-energy neutrinos and gravitational waves is possible with the current facilities, and probable with the planned experiments even for the moderate parameter sets. Non-detection of the coincident neutrino will be able to put meaningful constraints on these systems in the future.

\acknowledgment
This work partly made use of data supplied by the UK Swift Science Data Centre at the University of Leicester.

\bibliographystyle{ws-rv-van}
\bibliography{ssk}

%\printindex[aindx]                 % to print author index
%\printindex                         % to print subject index
\end{document}